\documentclass[journal]{IEEEtran}
\usepackage{qtree}
\usepackage{latexsym}
\usepackage{graphicx}

\def\PsfigVersion{1.9}
\ifx\undefined\psfig\else \fi

\let\LaTeXAtSign=\@
\let\@=\relax
\edef\psfigRestoreAt{\catcode`\@=\number\catcode`@\relax}
\catcode`\@=11\relax
\newwrite\@unused
\def\ps@typeout#1{{\let\protect\string\immediate\write\@unused{#1}}}
\ps@typeout{psfig/tex \PsfigVersion}

\def\figurepath{./}

\def\@nnil{\@nil}
\def\@empty{}
\def\@psdonoop#1\@@#2#3{}
\def\@psdo#1:=#2\do#3{\edef\@psdotmp{#2}\ifx\@psdotmp\@empty \else
    \expandafter\@psdoloop#2,\@nil,\@nil\@@#1{#3}\fi}
\def\@psdoloop#1,#2,#3\@@#4#5{\def#4{#1}\ifx #4\@nnil \else
       #5\def#4{#2}\ifx #4\@nnil \else#5\@ipsdoloop #3\@@#4{#5}\fi\fi}
\def\@ipsdoloop#1,#2\@@#3#4{\def#3{#1}\ifx #3\@nnil 
       \let\@nextwhile=\@psdonoop \else
      #4\relax\let\@nextwhile=\@ipsdoloop\fi\@nextwhile#2\@@#3{#4}}
\def\@tpsdo#1:=#2\do#3{\xdef\@psdotmp{#2}\ifx\@psdotmp\@empty \else
    \@tpsdoloop#2\@nil\@nil\@@#1{#3}\fi}
\def\@tpsdoloop#1#2\@@#3#4{\def#3{#1}\ifx #3\@nnil 
       \let\@nextwhile=\@psdonoop \else
      #4\relax\let\@nextwhile=\@tpsdoloop\fi\@nextwhile#2\@@#3{#4}}
\ifx\undefined\fbox
\newdimen\fboxrule
\newdimen\fboxsep
\newdimen\ps@tempdima
\newbox\ps@tempboxa
\long\def\fbox#1{\leavevmode\setbox\ps@tempboxa\hbox{#1}\ps@tempdima\fboxrule
    \advance\ps@tempdima \fboxsep \advance\ps@tempdima \dp\ps@tempboxa
   \hbox{\lower \ps@tempdima\hbox
  {\vbox{\hrule height \fboxrule
          \hbox{\vrule width \fboxrule \hskip\fboxsep
          \vbox{\vskip\fboxsep \box\ps@tempboxa\vskip\fboxsep}\hskip 
                 \fboxsep\vrule width \fboxrule}
                 \hrule height \fboxrule}}}}
\fi
\newread\ps@stream
\newif\ifnot@eof       
\newif\if@noisy        
\newif\if@atend        
\newif\if@psfile       
{\catcode`\%=12\global\gdef\epsf@start{
\def\epsf@PS{PS}
\def\epsf@getbb#1{%
\openin\ps@stream=#1
\ifeof\ps@stream\ps@typeout{Error, File #1 not found}\else
   {\not@eoftrue \chardef\other=12
    \def\do##1{\catcode`##1=\other}\dospecials \catcode`\ =10
    \loop
       \if@psfile
	  \read\ps@stream to \epsf@fileline
       \else{
	  \obeyspaces
          \read\ps@stream to \epsf@tmp\global\let\epsf@fileline\epsf@tmp}
       \fi
       \ifeof\ps@stream\not@eoffalse\else
       \if@psfile\else
       \expandafter\epsf@test\epsf@fileline:. \\%
       \fi
          \expandafter\epsf@aux\epsf@fileline:. \\%
       \fi
   \ifnot@eof\repeat
   }\closein\ps@stream\fi}%
\long\def\epsf@test#1#2#3:#4\\{\def\epsf@testit{#1#2}
			\ifx\epsf@testit\epsf@start\else
\ps@typeout{Warning! File does not start with `\epsf@start'.  It may not be a PostScript file.}
			\fi
			\@psfiletrue} 
{\catcode`\%=12\global\let\epsf@percent=
\long\def\epsf@aux#1#2:#3\\{\ifx#1\epsf@percent
   \def\epsf@testit{#2}\ifx\epsf@testit\epsf@bblit
	\@atendfalse
        \epsf@atend #3 . \\%
	\if@atend	
	   \if@verbose{
		\ps@typeout{psfig: found `(atend)'; continuing search}
	   }\fi
        \else
        \epsf@grab #3 . . . \\%
        \not@eoffalse
        \global\no@bbfalse
        \fi
   \fi\fi}%
\def\epsf@grab #1 #2 #3 #4 #5\\{%
   \global\def\epsf@llx{#1}\ifx\epsf@llx\empty
      \epsf@grab #2 #3 #4 #5 .\\\else
   \global\def\epsf@lly{#2}%
   \global\def\epsf@urx{#3}\global\def\epsf@ury{#4}\fi}%
\def\epsf@atendlit{(atend)} 
\def\epsf@atend #1 #2 #3\\{%
   \def\epsf@tmp{#1}\ifx\epsf@tmp\empty
      \epsf@atend #2 #3 .\\\else
   \ifx\epsf@tmp\epsf@atendlit\@atendtrue\fi\fi}

\chardef\psletter = 11 
\chardef\other = 12

\newif \ifdebug 
\newif\ifc@mpute 
\c@mputetrue 

\let\then = \relax
\def\r@dian{pt }
\let\r@dians = \r@dian
\let\dimensionless@nit = \r@dian
\let\dimensionless@nits = \dimensionless@nit
\def\internal@nit{sp }
\let\internal@nits = \internal@nit
\newif\ifstillc@nverging
\def \Mess@ge #1{\ifdebug \then \message {#1} \fi}

{ 
	\catcode `\@ = \psletter
	\gdef \nodimen {\expandafter \n@dimen \the \dimen}
	\gdef \term #1 #2 #3%
	       {\edef \t@ {\the #1}
		\edef \t@@ {\expandafter \n@dimen \the #2\r@dian}%
		\t@rm {\t@} {\t@@} {#3}%
	       }
	\gdef \t@rm #1 #2 #3%
	       {{%
		\count 0 = 0
		\dimen 0 = 1 \dimensionless@nit
		\dimen 2 = #2\relax
		\Mess@ge {Calculating term #1 of \nodimen 2}%
		\loop
		\ifnum	\count 0 < #1
		\then	\advance \count 0 by 1
			\Mess@ge {Iteration \the \count 0 \space}%
			\Multiply \dimen 0 by {\dimen 2}%
			\Mess@ge {After multiplication, term = \nodimen 0}%
			\Divide \dimen 0 by {\count 0}%
			\Mess@ge {After division, term = \nodimen 0}%
		\repeat
		\Mess@ge {Final value for term #1 of 
				\nodimen 2 \space is \nodimen 0}%
		\xdef \Term {#3 = \nodimen 0 \r@dians}%
		\aftergroup \Term
	       }}
	\catcode `\p = \other
	\catcode `\t = \other
	\gdef \n@dimen #1pt{#1} 
}

\def \Divide #1by #2{\divide #1 by #2} 

\def \Multiply #1by #2
       {{
	\count 0 = #1\relax
	\count 2 = #2\relax
	\count 4 = 65536
	\Mess@ge {Before scaling, count 0 = \the \count 0 \space and
			count 2 = \the \count 2}%
	\ifnum	\count 0 > 32767 
	\then	\divide \count 0 by 4
		\divide \count 4 by 4
	\else	\ifnum	\count 0 < -32767
		\then	\divide \count 0 by 4
			\divide \count 4 by 4
		\else
		\fi
	\fi
	\ifnum	\count 2 > 32767 
	\then	\divide \count 2 by 4
		\divide \count 4 by 4
	\else	\ifnum	\count 2 < -32767
		\then	\divide \count 2 by 4
			\divide \count 4 by 4
		\else
		\fi
	\fi
	\multiply \count 0 by \count 2
	\divide \count 0 by \count 4
	\xdef \product {#1 = \the \count 0 \internal@nits}%
	\aftergroup \product
       }}

\def\r@duce{\ifdim\dimen0 > 90\r@dian \then   
		\multiply\dimen0 by -1
		\advance\dimen0 by 180\r@dian
		\r@duce
	    \else \ifdim\dimen0 < -90\r@dian \then  
		\advance\dimen0 by 360\r@dian
		\r@duce
		\fi
	    \fi}

\def\Sine#1%
       {{%
	\dimen 0 = #1 \r@dian
	\r@duce
	\ifdim\dimen0 = -90\r@dian \then
	   \dimen4 = -1\r@dian
	   \c@mputefalse
	\fi
	\ifdim\dimen0 = 90\r@dian \then
	   \dimen4 = 1\r@dian
	   \c@mputefalse
	\fi
	\ifdim\dimen0 = 0\r@dian \then
	   \dimen4 = 0\r@dian
	   \c@mputefalse
	\fi
	\ifc@mpute \then
		\divide\dimen0 by 180
		\dimen0=3.141592654\dimen0
		\dimen 2 = 3.1415926535897963\r@dian 
		\divide\dimen 2 by 2 
		\Mess@ge {Sin: calculating Sin of \nodimen 0}%
		\count 0 = 1 
		\dimen 2 = 1 \r@dian 
		\dimen 4 = 0 \r@dian 
		\loop
			\ifnum	\dimen 2 = 0 
			\then	\stillc@nvergingfalse 
			\else	\stillc@nvergingtrue
			\fi
			\ifstillc@nverging 
			\then	\term {\count 0} {\dimen 0} {\dimen 2}%
				\advance \count 0 by 2
				\count 2 = \count 0
				\divide \count 2 by 2
				\ifodd	\count 2 
				\then	\advance \dimen 4 by \dimen 2
				\else	\advance \dimen 4 by -\dimen 2
				\fi
		\repeat
	\fi		
			\xdef \sine {\nodimen 4}%
       }}

\def\Cosine#1{\ifx\sine\UnDefined\edef\Savesine{\relax}\else
		             \edef\Savesine{\sine}\fi
	{\dimen0=#1\r@dian\advance\dimen0 by 90\r@dian
	 \Sine{\nodimen 0}
	 \xdef\cosine{\sine}
	 \xdef\sine{\Savesine}}}	      

\def\psdraft{
	\def\@psdraft{0}
}
\def\psfull{
	\def\@psdraft{100}
}

\psfull

\newif\if@scalefirst
\def\psscalefirst{\@scalefirsttrue}
\def\psrotatefirst{\@scalefirstfalse}
\psrotatefirst

\newif\if@draftbox
\def\psnodraftbox{
	\@draftboxfalse
}
\def\psdraftbox{
	\@draftboxtrue
}
\@draftboxtrue

\newif\if@prologfile
\newif\if@postlogfile
\def\pssilent{
	\@noisyfalse
}
\def\psnoisy{
	\@noisytrue
}
\psnoisy
\newif\if@bbllx
\newif\if@bblly
\newif\if@bburx
\newif\if@bbury
\newif\if@height
\newif\if@width
\newif\if@rheight
\newif\if@rwidth
\newif\if@angle
\newif\if@clip
\newif\if@verbose
\def\@p@@sclip#1{\@cliptrue}

\newif\if@decmpr

\def\@p@@sfigure#1{\def\@p@sfile{null}\def\@p@sbbfile{null}
	        \openin1=#1.bb
		\ifeof1\closein1
	        	\openin1=\figurepath#1.bb
			\ifeof1\closein1
			        \openin1=#1
				\ifeof1\closein1%
				       \openin1=\figurepath#1
					\ifeof1
					   \ps@typeout{Error, File #1 not found}
						\if@bbllx\if@bblly
				   		\if@bburx\if@bbury
			      				\def\@p@sfile{#1}%
			      				\def\@p@sbbfile{#1}%
							\@decmprfalse
				  	   	\fi\fi\fi\fi
					\else\closein1
				    		\def\@p@sfile{\figurepath#1}%
				    		\def\@p@sbbfile{\figurepath#1}%
						\@decmprfalse
	                       		\fi%
			 	\else\closein1%
					\def\@p@sfile{#1}
					\def\@p@sbbfile{#1}
					\@decmprfalse
			 	\fi
			\else
				\def\@p@sfile{\figurepath#1}
				\def\@p@sbbfile{\figurepath#1.bb}
				\@decmprtrue
			\fi
		\else
			\def\@p@sfile{#1}
			\def\@p@sbbfile{#1.bb}
			\@decmprtrue
		\fi}

\def\@p@@sfile#1{\@p@@sfigure{#1}}

\def\@p@@sbbllx#1{
		\@bbllxtrue
		\dimen100=#1
		\edef\@p@sbbllx{\number\dimen100}
}
\def\@p@@sbblly#1{
		\@bbllytrue
		\dimen100=#1
		\edef\@p@sbblly{\number\dimen100}
}
\def\@p@@sbburx#1{
		\@bburxtrue
		\dimen100=#1
		\edef\@p@sbburx{\number\dimen100}
}
\def\@p@@sbbury#1{
		\@bburytrue
		\dimen100=#1
		\edef\@p@sbbury{\number\dimen100}
}
\def\@p@@sheight#1{
		\@heighttrue
		\dimen100=#1
   		\edef\@p@sheight{\number\dimen100}
}
\def\@p@@swidth#1{
		\@widthtrue
		\dimen100=#1
		\edef\@p@swidth{\number\dimen100}
}
\def\@p@@srheight#1{
		\@rheighttrue
		\dimen100=#1
		\edef\@p@srheight{\number\dimen100}
}
\def\@p@@srwidth#1{
		\@rwidthtrue
		\dimen100=#1
		\edef\@p@srwidth{\number\dimen100}
}
\def\@p@@sangle#1{
		\@angletrue
		\edef\@p@sangle{#1} 
}
\def\@p@@ssilent#1{ 
		\@verbosefalse
}
\def\@p@@sprolog#1{\@prologfiletrue\def\@prologfileval{#1}}
\def\@p@@spostlog#1{\@postlogfiletrue\def\@postlogfileval{#1}}
\def\@cs@name#1{\csname #1\endcsname}
\def\@setparms#1=#2,{\@cs@name{@p@@s#1}{#2}}
\def\ps@init@parms{
		\@bbllxfalse \@bbllyfalse
		\@bburxfalse \@bburyfalse
		\@heightfalse \@widthfalse
		\@rheightfalse \@rwidthfalse
		\def\@p@sbbllx{}\def\@p@sbblly{}
		\def\@p@sbburx{}\def\@p@sbbury{}
		\def\@p@sheight{}\def\@p@swidth{}
		\def\@p@srheight{}\def\@p@srwidth{}
		\def\@p@sangle{0}
		\def\@p@sfile{} \def\@p@sbbfile{}
		\def\@p@scost{10}
		\def\@sc{}
		\@prologfilefalse
		\@postlogfilefalse
		\@clipfalse
		\if@noisy
			\@verbosetrue
		\else
			\@verbosefalse
		\fi
}
\def\parse@ps@parms#1{
	 	\@psdo\@psfiga:=#1\do
		   {\expandafter\@setparms\@psfiga,}}
\newif\ifno@bb
\def\bb@missing{
	\if@verbose{
		\ps@typeout{psfig: searching \@p@sbbfile \space  for bounding box}
	}\fi
	\no@bbtrue
	\epsf@getbb{\@p@sbbfile}
        \ifno@bb \else \bb@cull\epsf@llx\epsf@lly\epsf@urx\epsf@ury\fi
}	
\def\bb@cull#1#2#3#4{
	\dimen100=#1 bp\edef\@p@sbbllx{\number\dimen100}
	\dimen100=#2 bp\edef\@p@sbblly{\number\dimen100}
	\dimen100=#3 bp\edef\@p@sbburx{\number\dimen100}
	\dimen100=#4 bp\edef\@p@sbbury{\number\dimen100}
	\no@bbfalse
}
\newdimen\p@intvaluex
\newdimen\p@intvaluey
\def\rotate@#1#2{{\dimen0=#1 sp\dimen1=#2 sp
		  \global\p@intvaluex=\cosine\dimen0
		  \dimen3=\sine\dimen1
		  \global\advance\p@intvaluex by -\dimen3
		  \global\p@intvaluey=\sine\dimen0
		  \dimen3=\cosine\dimen1
		  \global\advance\p@intvaluey by \dimen3
		  }}
\def\compute@bb{
		\no@bbfalse
		\if@bbllx \else \no@bbtrue \fi
		\if@bblly \else \no@bbtrue \fi
		\if@bburx \else \no@bbtrue \fi
		\if@bbury \else \no@bbtrue \fi
		\ifno@bb \bb@missing \fi
		\ifno@bb \ps@typeout{FATAL ERROR: no bb supplied or found}
			\no-bb-error
		\fi
		\count203=\@p@sbburx
		\count204=\@p@sbbury
		\advance\count203 by -\@p@sbbllx
		\advance\count204 by -\@p@sbblly
		\edef\ps@bbw{\number\count203}
		\edef\ps@bbh{\number\count204}
		\if@angle 
			\Sine{\@p@sangle}\Cosine{\@p@sangle}
	        	{\dimen100=\maxdimen\xdef\r@p@sbbllx{\number\dimen100}
					    \xdef\r@p@sbblly{\number\dimen100}
			                    \xdef\r@p@sbburx{-\number\dimen100}
					    \xdef\r@p@sbbury{-\number\dimen100}}
                        \def\minmaxtest{
			   \ifnum\number\p@intvaluex<\r@p@sbbllx
			      \xdef\r@p@sbbllx{\number\p@intvaluex}\fi
			   \ifnum\number\p@intvaluex>\r@p@sbburx
			      \xdef\r@p@sbburx{\number\p@intvaluex}\fi
			   \ifnum\number\p@intvaluey<\r@p@sbblly
			      \xdef\r@p@sbblly{\number\p@intvaluey}\fi
			   \ifnum\number\p@intvaluey>\r@p@sbbury
			      \xdef\r@p@sbbury{\number\p@intvaluey}\fi
			   }
			\rotate@{\@p@sbbllx}{\@p@sbblly}
			\minmaxtest
			\rotate@{\@p@sbbllx}{\@p@sbbury}
			\minmaxtest
			\rotate@{\@p@sbburx}{\@p@sbblly}
			\minmaxtest
			\rotate@{\@p@sbburx}{\@p@sbbury}
			\minmaxtest
			\edef\@p@sbbllx{\r@p@sbbllx}\edef\@p@sbblly{\r@p@sbblly}
			\edef\@p@sbburx{\r@p@sbburx}\edef\@p@sbbury{\r@p@sbbury}
		\fi
		\count203=\@p@sbburx
		\count204=\@p@sbbury
		\advance\count203 by -\@p@sbbllx
		\advance\count204 by -\@p@sbblly
		\edef\@bbw{\number\count203}
		\edef\@bbh{\number\count204}
}
\def\in@hundreds#1#2#3{\count240=#2 \count241=#3
		     \count100=\count240	
		     \divide\count100 by \count241
		     \count101=\count100
		     \multiply\count101 by \count241
		     \advance\count240 by -\count101
		     \multiply\count240 by 10
		     \count101=\count240	
		     \divide\count101 by \count241
		     \count102=\count101
		     \multiply\count102 by \count241
		     \advance\count240 by -\count102
		     \multiply\count240 by 10
		     \count102=\count240	
		     \divide\count102 by \count241
		     \count200=#1\count205=0
		     \count201=\count200
			\multiply\count201 by \count100
		 	\advance\count205 by \count201
		     \count201=\count200
			\divide\count201 by 10
			\multiply\count201 by \count101
			\advance\count205 by \count201
		     \count201=\count200
			\divide\count201 by 100
			\multiply\count201 by \count102
			\advance\count205 by \count201
		     \edef\@result{\number\count205}
}
\def\compute@wfromh{
		\in@hundreds{\@p@sheight}{\@bbw}{\@bbh}
		\edef\@p@swidth{\@result}
}
\def\compute@hfromw{
	        \in@hundreds{\@p@swidth}{\@bbh}{\@bbw}
		\edef\@p@sheight{\@result}
}
\def\compute@handw{
		\if@height 
			\if@width
			\else
				\compute@wfromh
			\fi
		\else 
			\if@width
				\compute@hfromw
			\else
				\edef\@p@sheight{\@bbh}
				\edef\@p@swidth{\@bbw}
			\fi
		\fi
}
\def\compute@resv{
		\if@rheight \else \edef\@p@srheight{\@p@sheight} \fi
		\if@rwidth \else \edef\@p@srwidth{\@p@swidth} \fi
}
\def\compute@sizes{
	\compute@bb
	\if@scalefirst\if@angle
	\if@width
	   \in@hundreds{\@p@swidth}{\@bbw}{\ps@bbw}
	   \edef\@p@swidth{\@result}
	\fi
	\if@height
	   \in@hundreds{\@p@sheight}{\@bbh}{\ps@bbh}
	   \edef\@p@sheight{\@result}
	\fi
	\fi\fi
	\compute@handw
	\compute@resv}

\def\psfig#1{\vbox {
	\ps@init@parms
	\parse@ps@parms{#1}
	\compute@sizes
	\ifnum\@p@scost<\@psdraft{
		\special{ps::[begin] 	\@p@swidth \space \@p@sheight \space
				\@p@sbbllx \space \@p@sbblly \space
				\@p@sbburx \space \@p@sbbury \space
				startTexFig \space }
		\if@angle
			\special {ps:: \@p@sangle \space rotate \space} 
		\fi
		\if@clip{
			\if@verbose{
				\ps@typeout{(clip)}
			}\fi
			\special{ps:: doclip \space }
		}\fi
		\if@prologfile
		    \special{ps: plotfile \@prologfileval \space } \fi
		\if@decmpr{
			\if@verbose{
				\ps@typeout{psfig: including \@p@sfile.Z \space }
			}\fi
			\special{ps: plotfile "`zcat \@p@sfile.Z" \space }
		}\else{
			\if@verbose{
				\ps@typeout{psfig: including \@p@sfile \space }
			}\fi
			\special{ps: plotfile \@p@sfile \space }
		}\fi
		\if@postlogfile
		    \special{ps: plotfile \@postlogfileval \space } \fi
		\special{ps::[end] endTexFig \space }
		\vbox to \@p@srheight sp{
			\hbox to \@p@srwidth sp{
				\hss
			}
		\vss
		}
	}\else{
		\if@draftbox{		
			\hbox{\frame{\vbox to \@p@srheight sp{
			\vss
			\hbox to \@p@srwidth sp{ \hss \@p@sfile \hss }
			\vss
			}}}
		}\else{
			\vbox to \@p@srheight sp{
			\vss
			\hbox to \@p@srwidth sp{\hss}
			\vss
			}
		}\fi

	}\fi
}}
\psfigRestoreAt
\let\@=\LaTeXAtSign

\def\punto{\hspace*{\fill}\Box}

\date{}

\begin{document}

\title{An XML-based Multi-Agent System for Supporting Online Recruitment Services}

\author{Pasquale~De Meo,
        Giovanni Quattrone,
        Giorgio Terracina,
        and~Domenico~Ursino
\thanks{P. De Meo, G. Quattrone and D. Ursino are with DIMET (Dipartimento di Informatica, Matematica, Elettronica
e Trasporti) of University Mediterranea of Reggio Calabria, Reggio Calabria, Italy. G. Terracina is with the
Department of Mathematics of University of Calabria, Rende (CS), Italy. Email:
\{demeo,quattrone,ursino\}@unirc.it, terracina@mat.unical.it.}}


\markboth{IEEE Transactions on Systems, Man, and Cybernetics,~Vol.~37, No.~4,~2007}{An XML-based Multi-Agent
System for Supporting Online Recruitment Services}

\maketitle

\date{}

\begin{abstract}
In this paper we propose an XML-based multi-agent recommender system for supporting online recruitment
services. Our system is characterized by the following features: {\em (i)} it handles user profiles for
personalizing the job search over the Internet; {\em (ii)} it is based on the Intelligent Agent Technology;
{\em (iii)} it uses XML for guaranteeing a light, versatile and standard mechanism for information
representation, storing and exchange. The paper discusses the basic features of the proposed system, presents
the results of an experimental study we have carried out for evaluating its performance, and makes a
comparison between the proposed system and other e-recruitment systems already presented in the past.
\end{abstract}


\section{Introduction}
\label{sec:Introduction}

\PARstart{I}{n} the last decade Internet services emerged as a significant, both social and cultural,
phenomenon; in fact, presently, many organizations provide their customers with the possibility to access
their services also on the Internet.

In this scenario, {\em online recruitment services}, supporting both {\em individuals} looking for a job and
{\em companies} looking for employees, are assuming a prominent role. In such a context, generally, companies
populate a database of job proposals and individuals are supported in their job search by an engine based on
classical Information Retrieval (IR) techniques\footnote{See, for example, {\tt hotjobs.yahoo.com} or {\tt
http://www.jobfinder.com}.}. Several studies show that the number of users accessing these services is
dramatically increasing \cite{USBureau}; as a consequence, it is possible to foresee that, in the next years,
a huge amount of job proposals will be handled by means of the Internet. In such a situation, classical IR
techniques used by the present recruitment services could provide an individual with an excessively large
number of job proposals not interesting for him. This could result in a low user perceived quality of service
and, ultimately, in his decision to not access these services.

A way for solving this problem consists of realizing {\em personalized search engines}, which combine
classical IR techniques and user modeling methodologies \cite{LiYuMe04}. In the literature some approaches
based on {\em recommender systems} \cite{ReVa97} have been proposed to personalize search engines in various
application contexts, such as Web browsing \cite{Lieberman95}, e-commerce \cite{HeJeLe03}, e-learning
\cite{Zaiane02}, and so on. In addition, in some proposals, recommender systems have been combined with the
agent technology for making them {\em autonomous, reactive} and {\em pro-active} \cite{Wooldridge02}. This
paper provides a contribution in this setting; in fact, it proposes an XML-based multi-agent recommender
system that exploits rich user profiles to support personalized recruitment services.

Our system is based on the agent technology. In fact, a User Agent is associated with each user and manages
his profile, as well as the interaction with him; a Recruitment Agent supports each User Agent in the
selection of those job proposals appearing to be the most adequate for the corresponding user. The
exploitation of the intelligent agent technology allows our system to improve its autonomy and adaptiveness,
as well as to easily partition the various tasks among several entities, thus improving its scalability.
Finally, it allows an easy management of existing legacy systems by means of suitable wrappers; as a
consequence, our system can easily interact with pre-existing recruitment Web sites and this contributes to
enhance the completeness of obtained results.

Our system is XML-based. XML is a standard language for representing and exchanging information. It embodies
both representation capabilities, typical of HTML, and data management features, typical of DBMSs. XML
information bases are plain text documents and, consequently, they are light, versatile, easy to be exchanged
and can reside on different, both hardware and software, platforms. In spite of this simplicity, the
information representation rules embodied in this language are powerful enough to allow the management of
sophisticated queries.

As for the contribution of our paper in the field of e-recruitment systems, we observe that:

\begin{itemize}

\item In order to select the job proposals that best satisfy user requirements, a traditional e-recruitment
system considers only the query of the user (that specifies his preferences and constraints), along with the
features of available job proposals. On the contrary, our system takes into account not only the query of the
user but also his reaction to previous job proposal recommendations it made to him.

Moreover, it analyzes similitudes and/or correlations among job proposals to produce its suggestions; in this
way, it can highlight some proposals that are usually ignored by traditional e-recruitment systems and that
might be appealing for the user.

\item In a traditional e-recruitment system, a user can query the corresponding (internal) database. As a
consequence, if the system's answer to a query is insufficient, the user must access other e-recruitment
systems and re-submit the query; after this, he must compose the results provided by each system to construct
his final list of job proposals. This activity might be time consuming, since the user is obliged to contact
and query several e-recruitment systems. In addition, it might be prone to errors, since two e-recruitment
systems might exploit different methodologies and use different terminologies to both select and represent
job proposals. Specifically, as for methodologies, it could happen that the same job proposal for a user
$u_i$ receives contrasting scores by different e-recruitment systems. As a consequence, when $u_i$ compares
the results provided by the various systems, he might be confused and, therefore, incapable of making a
decision about its suitability for him. As for terminologies, two independent e-recruitment systems might
present to a user the same job proposal by exploiting different terms. As an example, a job proposal
regarding the design, the development and the validation of software programs might be called ``Program
Application Support Specialist'' by an e-recruitment system, and ``Software Test and Software Design
Engineer'' by another one. In situations like that previously described, since the two systems are totally
independent each other, there is no form of coordination and, consequently, no way to inform a user that two
or more terms represent the same job proposal.

\end{itemize}

Our system is capable of facing all these problems; specifically, as it will be clarified in the following,
it continuously monitors existing online recruitment services and collects job proposals coming from them; as
a consequence, user search is not restricted to the database associated with an e-recruitment system but it
spreads across disparate sources. This system organization allows a user to pose his queries in a {\em
transparent} way; in other words, in order to obtain an answer, he could be unaware of the models and the
languages of the involved databases, as well as of their internal structure.

As for the Recommender Systems research area, the system we propose in this paper provides the following
novelties:

\begin{itemize}

\item A traditional Recommender System usually identifies the ``Top K'' recommendations, i.e., the $K$
information items (in our case, job proposals) having the highest relevance for the user. The coefficient $K$
is usually constant and pre-defined. In line with current advances \cite{Lawrence*01}, our system can analyze
user reactions to past proposals to assess if the number of job proposals it presents to users is too low
(resp., too high) and, in the affirmative case, it can dynamically augment (resp., reduce) this number.

\item In content-based Recommender Systems, a user profile is generally represented as a vector of pairs
$\langle k_i, w_i \rangle$, where $k_i$ is a keyword and $w_i$ is a weight denoting the importance of $k_i$
for the user. On the contrary, in our system, user profiles are richer and take into account various
characteristics and preferences of involved users. Even if this implies a higher complexity in their
management, the improvement of results that can be obtained is very satisfying, as it will be clarified in
Section \ref{sub:comparison}.

\item Our approach applies in the Recommender Systems field some mathematic tools generally used in other
application contexts, such as {\em manufacturing engineering} \cite{GiBaLi02}, {\em bioinformatics}
\cite{SiSaSa03} and  {\em econometrics} \cite{Yikang98}. Interestingly enough, the computational effort it
requires is not particularly heavy, as it will be demonstrated in Section \ref{sec:personalization}.

\end{itemize}

\section{General Architecture of our system}
\label{sec:Overview}

Our system consists of two main agent typologies, namely: {\em (i)} a {\em User Agent} ($UA$), that manages
the profile of a user, as well as the interaction with him, and {\em (ii)} a {\em Recruitment Agent} ($RA$),
that handles both job proposals and recommendation activities.

Moreover, a Job Proposal Database ($JPD$) is used to store all available job proposals. $JPD$ can be
populated by companies who can insert, remove or modify job proposals by means of a {\em Company Agent},
i.e., a suitable Interface Agent, analogous to that described in \cite{CeDa96}. In addition, $JPD$ can be
automatically enriched by suitable {\em Wrapper Agents}, analogous to that described in \cite{Hsinchun*98},
that continuously monitor existing online recruitment services, find new job proposals and store them in
$JPD$, if they are not already present. $JPD$ can be described by an XML document; the corresponding XML
Schema is shown in the Appendix available at the address {\footnotesize {\tt
http://www.ing.unirc.it/ursino/tsmc05/Appendix.pdf}}.

Specifically, $JPD$ stores a set of job proposals; a job proposal $JP_l$ is described by a tuple $\langle
JID_l, JURL_l, JTopicSet_l, JCharacteristicSet_l \rangle$, where:

\begin{itemize}

\item $JID_l$ is an identifier;

\item $JURL_l$ is the url where the complete description of the job proposal is available;

\item $JTopicSet_l = \{ JTopic_{l_1},JTopic_{l_2},\ldots, JTopic_{l_m} \}$ is a set of topics describing
$JP_l$; a topic is characterized by its name;

\item $JCharacteristicSet_l$ is a dynamic set of {\em characteristics} associated with $JP_l$. Each
characteristic is described by a pair $\langle Feature, Value \rangle$; some of the possible features are the
salary associated with the job proposal, the city/country of the job proposal, the foreign languages, the
skills, the years of experience and the academic title(s) required from the candidate.

\end{itemize}

\begin{figure}[t]
 \centerline{\psfig{figure=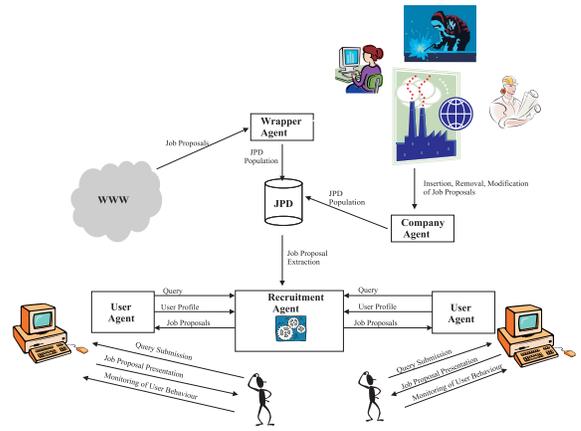,width=7.5cm}}
 \caption{General architecture of our system}
 \label{fig:architecture}
\end{figure}

Observe that we have chosen to store available job proposals into an independent knowledge base and not in
the ontology of a specific agent; in this way, several independent and specialized agents can operate in our
system to automatically and continuously update and refine them.

The general architecture of our system is shown in Figure \ref{fig:architecture}. Its behaviour is as
follows. When a user $u_i$ wants to perform a job search, he submits a query to the associated User Agent
$UA_i$. After this, $UA_i$ contacts $RA$ and provides it with both the query and the profile of $u_i$. $RA$
selects from $JPD$ the available job proposals satisfying user query, ranks them on the basis of the presumed
user preferences and selects those ones that best fit the past interests of $u_i$, as well as other
interests, someway related to those $u_i$ considered in the past and that he still disregarded. After this,
$RA$ sends the selected proposals to $UA_i$ that presents them to $u_i$; $UA_i$ monitors $u_i$ and, when he
performs his choices, it updates his profile accordingly.

As previously pointed out, the role of XML in our system is crucial. In fact:

\begin{itemize}

\item Agent ontologies are stored as XML documents; as a consequence, they are light, versatile, easy to be
exchanged and can reside on different hardware and software platforms. In spite of this simplicity, the
information representation rules embodied in XML are powerful enough to allow a sophisticated information
management.

\item The agent communication language adopted in our system is ACML \cite{GrLa99}; this is the XML encoding
of the standard Agent Communication Language.

\item The extraction of information from the various data structures is carried out by means of XQuery
\cite{XQuery}. This is becoming the standard query language for the XML environment. Since XQuery is based on
the XML framework, it can handle a large variety of data. It has capabilities typical of database query
languages as well as features typical of document management systems.

\item The manipulation of agent ontologies is performed by means of the Document Object Model (DOM)
\cite{DOM}. DOM makes it possible for programmers to write applications working properly on all browsers and
servers as well as on a large variety of both hardware and software platforms.

\end{itemize}

The architecture of our system can be considered {\em mixed}, i.e., partially centralized and partially
distributed. In this sense it follows the ideas expressed in \cite{HaJeMo05}, where an architecture for
handling telecommunication networks, based on the presence of an auctioneer agent that carries out most of
the negotiation activities, is described. It can be considered centralized for the presence of the
Recruitment Agent that performs most of the activities concerning the selection of the job proposals best
fitting user exigencies. On the other hand, it can be assumed to be distributed for the presence of many User
Agents, Wrapper Agents and Company Agents that continuously cooperate with the Recruitment Agent for
performing the whole recruitment task.

Observe that, in our system, there exists only a {\em central} Recruitment Agent. Such a choice is justified
by the following motivations:

\begin{itemize}

\item {\em Network Congestion}. In our architecture, each User Agent is in charge of forwarding both the User
Profile and the user query to the Recruitment Agent that processes this query and sends the corresponding
results to it; as a consequence, the two agents exchange a small number of (generally simple) messages and
this prevents network overload. In a distributed implementation of the Recruitment Agent activities, various
Recruitment Agents would cooperate for processing user queries. As a consequence, they should continuously
communicate and exchange messages; this would generate a high network traffic and the whole system would
quickly be overwhelmed with messages among agents. This problem is particularly relevant in our reference
scenario where the number of job seekers, and consequently of User Agents, is very high; this would cause an
overwhelming amount of exchanged messages and, ultimately, significant latencies and users dissatisfaction.

\item {\em Task Assignment}. In a distributed implementation of the Recruitment Agent activities, a query is
propagated through several Recruitment Agents that spend memory and CPU resources to process it. In this
scenario, it could happen that some Recruitment Agents would be overwhelmed with requests, whereas other ones
would be almost unoccupied. In order to optimize the system performances, it would be necessary to define
protocols for intelligently assigning tasks to the various Recruitment Agents. These protocols appear very
complex to be realized because the potential job seekers (and, consequently, the potential User Agents) might
be very numerous and geographically sparse.

\item {\em Message Duplication}. In a distributed implementation of the Recruitment Agent activities,
multiple copies of a query might be received by the same Recruitment Agent (we call these copies {\em
duplicated messages}). To better clarify this concept and its consequences, consider four Recruitment Agents
$RA_{d_1}$, $RA_{d_2}$, $RA_{d_3}$ and $RA_{d_4}$, and assume that: {\em (i)} a user submits a query
$\overline{Q}$ to $RA_{d_1}$; {\em (ii)} $RA_{d_1}$ is not capable of answering $\overline{Q}$; {\em (iii)}
$RA_{d_1}$ asks $RA_{d_2}$ and $RA_{d_3}$ to answer $\overline{Q}$; {\em (iv)} $RA_{d_2}$ and $RA_{d_3}$ are
only capable of producing a partial answer to $\overline{Q}$ (e.g., the number of job proposals retrieved by
them is lesser than that required by the user); {\em (v)} $RA_{d_2}$ and $RA_{d_3}$ might {\em independently}
forward $\overline{Q}$ to $RA_{d_4}$ that would receive and, in its turn, could process and possibly transmit
$\overline{Q}$ twice. This example clearly shows that duplicated messages are pure overhead; they increase
the network congestion and determine a waste of CPU and memory resources for the Recruitment Agents that are
often obliged to process the same query more than once. In spite of the resource waste caused by them, they
do not improve system accuracy since they do not contribute to increase the chance of finding job proposals
relevant for user.

These problems are emphasized in an e-recruitment scenario, where the number of job seekers contemporarily
accessing our system could be very high; in this case, the probability of duplicating messages might be
extremely high. As a consequence, suitable mechanisms for identifying duplicated messages and avoiding to
process and forward them are required; however, these mechanisms are generally difficult to be designed and
implemented.

\end{itemize}

Interestingly enough, these motivations are in accordance also with the ideas illustrated in \cite{DaPaJe03}.

Finally, we would like to point out that our system does not provide for any additional layer between the
User Agent and the Recruitment Agent. This choice is justified by the following motivations:

\begin{itemize}

\item {\em Ability to cooperate with other e-recruitment systems}. One of the most important capabilities of
our system consists of its ability to effectively cooperate with other e-recruitment systems (see below). A
layered architecture does not appear particularly suitable for pursuing such an objective. In fact, in a
layered architecture, system functionalities are associated with different abstraction levels, each
represented by a layer; the abstraction degree generally increases from the lowest levels to the highest
ones. The presence of different abstraction levels has the following consequences:

\begin{itemize}

\item The interaction between two layered systems requires each of them to have a deep knowledge of the
architectural features of the other one. In fact, each system should know the abstraction level associated
with each layer of the other system to properly exchange information and cooperate. Such a knowledge is often
missing and, hence, the mapping from a layer in a system to a layer in another one might be not obvious and,
in some cases, not possible.

\item Many systems cannot be easily structured in a layered fashion. As a consequence, if our system would be
layered, we could have a scenario in which layered and non-layered architectures coexist; in this case,
suitable communication protocols between layered and non-layered systems are compulsory; their realization,
however, is often a hard and expensive activity.

\end{itemize}

\item {\em Performance Issues}. Several studies point out that the number of users accessing e-recruitment
systems is rapidly increasing \cite{USBureau}. As a consequence, an e-recruitment system should be capable of
quickly processing user queries. Such a requirement cannot be easily satisfied in a layered system; in fact,
in this case, each layer supplies services to the layer above it and operates as a client for the layer below
it. Therefore, in a layered system, a continuous message exchange between low-level layers and the high-level
ones is required. Since a message often has to pass through many layers, the overhead associated with message
exchange is often significant and might determine a performance decay. Various techniques have been proposed
to boost the performances of a layered system (e.g., it is possible to couple non-adjacent layers) but they
seem too difficult to be realized in practical contexts, especially in our application case where the various
layers could belong to different owners.

\item {\em Reliability Issues}. The fast identification of faults and the quick reaction to them cannot be
easily performed in a layered architecture since a failure in a layer might determine the crash of the whole
system. Even the (possibly non fast) fault identification might be difficult to be carried out since the
functionalities of the low-level layers are often hidden to the high-level ones and, consequently, an
application running on high-level layers has many difficulties in identifying where a certain problem has
occurred.

\end{itemize}

In the following sections we provide a detailed description of the two main agents involved in our system,
namely the User Agent and the Recruitment Agent.

\section{The User Agent}
\label{sec:UseAgent}

\subsection{Ontology}
\label{subsec:uaontology}

The ontology of the User Agent $UA_i$, associated with a user $u_i$, stores the profile of $u_i$. It
contains:


\begin{itemize}

\item The code $UID_i$ identifying $u_i$.

\item A set $TopicSet_i$ storing the preferences of $u_i$; each preference corresponds to a topic which $u_i$
looked for in the past. Each element $Topic_{i_j}$ of $TopicSet_i$ is represented by a tuple $\langle
TopicName_{i_j}, Count_{i_j}, First\-TimeStamp_{i_j} \rangle$, where: {\em (i)} $TopicName_{i_j}$ indicates
the name of $Topic_{i_j}$; {\em (ii)} $Count_{i_j}$ is an {\em access counter} denoting how many times
$Topic_{i_j}$ has been specified in a query submitted by $u_i$; {\em (iii)} $FirstTimeStamp_{i_j}$ indicates
the first time instant in which $Topic_{i_j}$ has been specified in a query of $u_i$. As it will be clarified
in the following, these last two coefficients allow the relevance of $Topic_{i_j}$ for $u_i$ to be measured.

\item A dynamic set $ConstraintSet_i$ of constraints that $u_i$ fixes for the desired job. Each constraint is
described by a pair $\langle Feature, Value \rangle$; some of the possible constraints are the minimum salary
$u_i$ would like to earn, the city/country where he desires to work, the foreign language(s) he knows, his
skills, the years of experience he has attained, and his academic title(s).

\item A set $PastQueries_i$, storing information about the queries that $u_i$ posed to the system in the
past. Each element of $PastQueries_i$ consists of a pair $\langle \sigma_i^k, \alpha_i^k \rangle$.

$\sigma_i^k$ is the {\em satisfaction} coefficient; it belongs to the real interval $[0,1]$ and indicates how
much $u_i$ appreciated the recommendations provided by the system as a response to his $k^{th}$ query. In the
literature many proposals have been presented to measure the satisfaction of a user for a set of
recommendations (see, for example, \cite{Carreira*04,ThRa04}). In this paper we follow the ideas illustrated
in \cite{ThRa04} and define $\sigma_i^k$ as follows. Let $JPList_i^k$ be the set of job proposals recommended
by the system as a response to the $k^{th}$ query of $u_i$ and let $AcceptedJP_i^k$ be the set of job
proposals accepted by $u_i$; $\sigma_i^k$ can be defined as $\sigma_i^k =
\frac{|AcceptedJP_i^k|}{|JPList_i^k|}$.

$\alpha_i^k$ is the {\em audacity} coefficient; it belongs to the real interval $[0,1]$; the higher
$\alpha_i^k$ is, the higher the number of job proposals returned by the system as its answer to the query
$Q_i^k$ will be. $\alpha_i^k$ is computed by the Recruitment Agent; the various strategies for its
computation are shown in Section \ref{sec:personalization}; they have been designed in such a way to make our
system adaptive against the past behaviour of $u_i$.

\end{itemize}

The XML Schema associated with the ontology of $UA_i$ is shown in the Appendix.

\subsection{Behaviour}
\label{subsec:uabehavior}

The behaviour of $UA_i$ consists of the following steps.

\begin{itemize}

\item {\em Step 1.} When $u_i$ wants to perform a job search, $UA_i$ supports him in the construction of a
corresponding query $Q_i^k$ by means of a suitable wizard. $Q_i^k$ consists of a pair $ \langle
SelDegree_i^k,$ $QTSet_i^k \rangle$. $SelDegree_i^k$, belonging to the real interval $[0,1]$, is the {\em
Selectivity Degree}; it indicates how much the system should be selective in the search of proposals (see
below); the wizard associated with $UA_i$ proposes to $u_i$ some possibilities and he chooses the most
suitable one for him in a friendly, guided manner. $QTSet_i^k = \{QTopic_{i_1}^k, \ldots, QTopic_{i_q}^k\}$
is the set of topics describing the job proposals currently interesting for $u_i$.

Moreover, if $u_i$ wants to update the constraints for a desired job, $UA_i$ provides him with a suitable
graphical interface that allows him to view the constraints he has defined, as well as to add, drop or modify
some of them.

\item {\em Step 2.} $UA_i$ updates $TopicSet_i$; this task is performed by inserting in it those topics of
$Q_i^k$ not already present therein and by increasing the access counters of the topics corresponding to
$QTopic_{i_1}^k, \ldots, QTopic_{i_q}^k$.

\item {\em Step 3.} $UA_i$ contacts $RA$ and supplies it both $Q_i^k$ and the profile of $u_i$. $RA$ selects
the job proposals answering $Q_i^k$ that best fit the past interests of $u_i$, as well as other interests,
someway related to those $u_i$ considered in the past and that he still disregarded. Then, it sends to $UA_i$
both the selected job proposals and the audacity coefficient $\alpha_i^k$ it used.

\item {\em Step 4.} $UA_i$ presents to $u_i$ the job proposals provided by $RA$. $u_i$ accepts those ones
appearing to be the closest to his exigencies and rejects the other ones. After this, $UA_i$ computes the
satisfaction degree $\sigma_i^k$ that $u_i$ showed for the provided recommendations, and stores the pair
$\langle \sigma_i^k, \alpha_i^k \rangle$ into $PastQueries_i$.

\item {\em Step 5.} In order to avoid an excessive growth of $TopicSet_i$, $UA_i$ executes a pruning activity
on it. Such a task is carried out by computing the relevance for $u_i$ of each topic $Topic_{i_j}$ stored in
$TopicSet_i$ and by removing from $TopicSet_i$ those topics presenting a relevance smaller than a certain
threshold. The {\em relevance} of $Topic_{i_j}$ at the time instant $T$ is computed by means of the formula
$r(Topic_{i_j},T) = \frac{Count_{i_j}}{T-FirstTimeStamp_{i_j}}$. This formula is justified by considering
that the more a user is interested in a topic, the more frequently he asks information about it.

\end{itemize}

In order to show the exploitation of ACML and XQuery in our system, in the Appendix, we present the ACML
message that $UA_i$ sends to $RA$ for requiring a set of recommendations and the query that $UA_i$ executes
for identifying those topics of $Q_i^k$ not already present in $TopicSet_i$.

\subsection{Influences of job market specifics in the design of the User Agent architecture}
\label{sub:influences}

The design of our User Agent model has been influenced by various job market specifics. In the following we
examine the most relevant of them.

\subsubsection{Construction of a network of job seekers}
\label{sub:network-job-seekers}

Several studies point out the importance of constructing a job seeker network (JSN), i.e., a community of job
seekers that share and exchange information about their past job experiences. A JSN is an effective channel
for disseminating job information and can successfully support its users in their job search: as an example,
a study presented in \cite{WaZe04} shows that the probability of a user to find a job grows with the increase
of the size of the network he belongs to.

Actually, the construction of a $JSN$ is a delicate activity. In fact, job seekers, due to both selfishness
and rivalry, might be unwilling to cooperate. To overcome this drawback, it is necessary to create a
free-of-competition environment based on trustful relationships among its members. In order to carry out this
task, it is necessary to enhance the cooperation among job seekers belonging to the same job context for whom
a rivalry might not exist; think, for example, to job seekers having different experience degrees (e.g., a
senior software programmer who helps a newly graduated person to find his first job).

The experience shows also that information sharing among members of a JSN increases their capability of
posing precise queries. In fact, users accessing e-recruitment services are extremely heterogeneous: job
seekers having a deep knowledge of a job domain often coexist with other seekers having a superficial
knowledge of it. An expert is capable of formulating a query that really captures his needs; on the contrary,
a novice might ignore the information content of the database storing available job proposals and,
consequently, might be unable to formulate precise queries. As previously pointed out, a JSN might support
novices to compose their queries; specifically, a novice might ask the experts of a JSN to support him in his
query composition.

Our User Agent model has been designed in such a way to facilitate the creation of a JSN; specifically, a
User Agent $UA_i$, on behalf of $u_i$, could contact other User Agents for finding job seekers having a
profile similar to $u_i$ but being more expert than him; such a task can be carried out by taking into
account the lists of topics specified in user profiles and by comparing them by means of the Jaccard
Coefficient \cite{HaKa01}. When the list of job seekers similar to $u_i$ is available, $u_i$ can ask $UA_i$
to activate a contact with one or more of them in such a way to require a support to formulate his queries.

\subsubsection{Integration with e-learning systems}
\label{sub:integration-learning-systems}

Several studies point out that the capability of a job seeker to continuously update his know-how has a
significant implication on the success of his job search, as well as on the advancement of his career; as a
consequence, learning activities should be regarded as a lifelong process. For this reason a strict
integration between e-recruitment and e-learning systems might provide a job seeker with a significant
competitive advantage.

A key component of most e-learning systems is the so-called student profile, storing information about the
student background knowledge and information needs. In this way, it is possible to diversify the learning
process on the basis of user skills and necessities.

Our User Agent model can make the interaction between our system and an e-learning one easier; specifically,
a User Agent could transfer the corresponding User Profile to an e-learning system that might construct an
initial student profile starting from it. Conversely, the profile of a student that acquired some skills with
the support of an e-learning system could be exploited to construct an initial user profile for the User
Agent of our e-recruitment system.

\subsubsection{Efficient retrieval of curricula}
\label{sub:curricula-retrieval}

Presently, a large number of curricula vitae is available on the Internet. Generally, a curriculum consists
of a title and a set of additional information provided by the job seeker (e.g., his past experience). These
curricula can be consulted by businesses; a common way to do this activity is the so called ``curricula
spidering'', i.e., the exploitation of a software agent that tracks down curricula and finds those of
interest for businesses.

The search of curricula is generally performed by means of a traditional search engine; in this case the
business manager specifies some words to the engine and this last searches for those curricula containing
these words. These techniques might lead to irrelevant and spurious results; in fact, a curriculum might be
too vague; it might be not updated; last, but not the least, the words appearing in a curriculum might not
belong to the vocabulary of terms used by business managers in their searches.

Our User Agent model is capable of enhancing the curricula search capabilities of businesses; in fact:

\begin{itemize}

\item A User Agent is in charge of monitoring the behaviour of its user and this allows it to maintain the
corresponding profile always updated.

\item A User Agent learns the profile of its user on the basis of the job proposals that he searches for and,
eventually, accepts or rejects. In this way, the topics specified in a User Profile derive from the job
proposal features directly inserted by businesses; such a policy avoids ambiguity, that could arise when a
term present in a traditional curriculum could be interpreted in more than one way by a business manager, and
protects the job seeker against his lack of knowledge of the terms that are currently used by businesses.

\end{itemize}

\section{The Recruitment Agent}
\label{sec:RecommenderAgent}

\subsection{Ontology}\label{sub:RAOntology}
The ontology of a Recruitment Agent $RA$ consists of a set of job proposals; as pointed out in Section
\ref{sec:Overview}, a Job Proposal $JP_l$ is described by a tuple $\langle JID_l, JURL_l, JTopicSet_l,
JCharacter\-isticSet_l \rangle$.

The XML Schema associated with the ontology of $RA$ is shown in the Appendix.

\subsection{Behaviour}\label{sub:RABehaviour}

$RA$ is activated by $UA_i$ when $u_i$ wants to perform a query $Q_i^k$ for searching job proposals. It
receives $Q_i^k$ and the profile of $u_i$; it returns to $UA_i$: {\em (i)} the job proposals answering
$Q_i^k$ that best fit the past interests of $u_i$, as well as other interests, someway related to those $u_i$
considered in the past and that he still disregarded; {\em (ii)} the audacity coefficient $\alpha_i^k$ it
used to select the job proposals. In order to perform its activity, $RA$ carries out the following steps:

\begin{itemize}

\item {\em Step 1.} It queries $JPD$ for retrieving the job proposals matching $Q_i^k$ ({\em keyword-based
filtering}) and satisfying user constraints ({\em feature-based filtering}); to this purpose it applies
classical Information Retrieval techniques \cite{BaRi99}.

\item {\em Step 2.} It ranks the selected proposals on the basis of user preferences, as stored in
$TopicSet_i$. Such a task is carried out as follows: let $JP_l = \langle JID_l, JURL_l, JTopicSet_l, $
$JCharacter\-isticSet_l \rangle$ be a job proposal and let $TS_{il} = \{ Topic_{i_j} \mid Topic_{i_j} \in
TopicSet_i,  TopicName_{i_j} \in JTopicSet_l \}$ be the set of topics of $JP_l$ appearing to be interesting
for $u_i$. $RA$ assigns an interest degree to $JP_l$, for the user $u_i$ at the current time instant $T$,
according to the formula $\rho(TS_{il}, T)= \sum\limits_{Topic_{i_j} \in TS_{il}} \ r(Topic_{i_j},T)$. Here,
the relevance $r$ has been defined in Section \ref{subsec:uabehavior}. Clearly, the higher the value of
$\rho(TS_{il}, T)$ is, the higher the relevance of $JP_l$ for $u_i$ will be. At the end of this phase, $RA$
constructs $JPTempList_i^k$, obtained by sorting the selected job proposals on the basis of their interest
degrees.

Observe that a more complex formula could be adopted for $\rho$. However, in this way, the complexity of the
Recruitment Agent would unnecessarily increase. In fact, since, during the selection of job proposals, our
approach takes into account also user past preferences and satisfaction, a simple ranking function is
sufficient. This is confirmed also by the experimental results shown in Section \ref{sec:Experimental}.

$JPTempList_i^k$ is already a good candidate to be presented to $u_i$. However, two further improvements
could be performed by $RA$ to make it more adequate. Firstly, we observe that $JPTempList_i^k$ ranks the job
proposals on the basis of their relevance for $u_i$, according only to his {\em past preferences}; however,
it does not consider topics, someway related to those $u_i$ considered in the past and that he still
disregarded. Secondly, $JPTempList_i^k$ could still contain quite a large number of proposals. The next steps
performed by $RA$ are devoted to carry out these improvements.

\item {\em Step 3.} $RA$ constructs the set $SeedProposals_i^k$ obtained by selecting the first $\lceil
SelDegree_i^k  \times |JPTempList_i^k| \rceil$ proposals of $JPTempList_i^k$. The elements of
$SeedProposals_i^k$ are used by $RA$ as seeds to choose other job proposals containing topics, someway
related to those $u_i$ considered in the past and that he still disregarded.

\item {\em Step 4.} $RA$ constructs the final list $JPList_i^k$ of recommended job proposals as $JPList_i^k =
SeedProposals_i^k \cup \{ JP_l \mid JP_l \in JPTempList_i^k, JP_m \in SeedProposals_i^k, \delta(JP_l, JP_m)
\leq \alpha_i^k \}$. Here $ \delta (JP_{l},JP_{m}) $ measures the ``dissimilarity degree'' between $JP_l$ and
$JP_m$. It is defined as $ \delta (JP_{l},JP_{m}) = 1 - \frac {2 |JTopicSet_{l} \cap JTopicSet_{m} | }
{|JTopicSet_{l}| + |JTopicSet_{m}|}$, where $JTopicSet_l$ and $JTopicSet_m$ represent the sets of topics
associated with $JP_l$ and $JP_m$, respectively, and the symbol $|S|$ represents the cardinality of a set
$S$. Interestingly enough, the value $\frac{2 |S_1 \cap S_2|}{|S_1|+|S_2|}$ is known as Dice's coefficient in
the literature. It is easy to show that $\delta$ ranges between 0 (if the two job proposals coincide) and $1$
(if the two job proposals are completely different).

The audacity coefficient $\alpha_i^k$ is dynamically computed by $RA$ on the basis of the feedback that $u_i$
showed for past recommendations. In this paper we propose three different strategies for the computation of
this coefficient; they are extensively described in Section \ref{sec:personalization}.

$JPList_i^k$, obtained at the end of this step, contains at least the seed proposals; moreover, it could
include also some job proposals relative to topics not particularly relevant for $u_i$ in the past, but that
could be of interest for him in the future (since they are sufficiently similar to those he judged relevant
in the past). In the selection of this last category of proposals $\delta$ plays a key role; in fact, it
measures the dissimilarity degree of two job proposals on the basis of their topics, i.e., of their meaning
and semantics, without considering the relevance that $u_i$ assigned to them in the past.

\item {\em Step 5.} $RA$ sends $JPList_i^k$ and $\alpha_i^k$ to $UA_i$.

\end{itemize}

\subsubsection{Strategies for the audacity computation}
\label{sec:personalization}

In our system we have chosen to dynamically update the audacity coefficient after each query performed by the
user $u_i$. This allows it to be very sensitive to the user judgements about its recommendations and
significantly improves its accuracy. However, this makes the manual update of the audacity coefficient a very
difficult task. In fact, in our opinion, asking the user to manually update the audacity coefficient after
each query is obtrusive. In addition, even with the exploitation of suitable graphic interfaces, the user
could be not capable of understanding the role of the audacity coefficient and to choose the most correct
value for it.

However, in our prototype, in order to consider the possible presence of very smart users, we have inserted a
module allowing a smart user to personally modify the values of the audacity coefficient determined by the
system.

In this section we illustrate three different strategies that may be used for automatically computing the
audacity coefficient $\alpha_i^k$ that $RA$ associates with $Q_i^k$. Recall that the higher this coefficient
is, the higher the number of job proposals returned by the system for answering $Q_i^k$ will be; this
strictly depends on the satisfaction that $u_i$ showed for the recommendations of $RA$ in the past. As a
consequence, any strategy for the computation of $\alpha_i^k$ must evaluate the satisfaction of $u_i$ w.r.t.
the previous recommendations of $RA$. As previously pointed out, the choice to make audacity dependent on
satisfaction allows our system to be adaptive against the past behaviour of $u_i$.

\paragraph{Strategy 1} {\em Positive and Negative Feedback} (PNF).

In this strategy, $\alpha_i^k$ is dynamically updated on the basis of the feedback provided by $u_i$ for the
last recommendation; this is encoded in the satisfaction coefficient $\sigma_i^{k-1}$ relative to the
recommendations associated with the query $Q_i^{k-1}$. Recall that $\sigma_i^{k-1}$ is the fraction of job
proposals accepted by $u_i$ w.r.t. those recommended by the system when it answered $Q_i^{k-1}$. PNF works as
follows:

\begin{itemize}

\item If, during the execution of $Q_i^{k-1}$, the number of accepted recommendations has been greater than
the number of the rejected ones (i.e., $\sigma_i^{k-1} > \frac{1}{2}$), then it is possible to argue that
$u_i$ has appreciated the system recommendations, since he accepted most of them. As a consequence, if the
system would send further recommendations, it is extremely probable that $u_i$ would receive other proposals
interesting for him. Now, since in our system the amount of proposals sent to the user as the answer to the
query $Q_i^k$ is strictly related to the audacity coefficient $\alpha_i^k$, increasing this coefficient would
imply an increase of the number of proposals recommended by the system when answering $Q_i^k$.

Clearly, the key issue in this reasoning is the correct estimation of the increase of $\alpha_i^k$. In fact,
if such an increase is too high, then $u_i$ would receive several useless proposals and his perceived quality
of the overall recommendations could decrease. On the contrary, if the increase is too low, the system could
miss to recommend some proposals interesting for $u_i$. The previous reasoning led us to define the increase
$\varepsilon_i^k$ to apply on the audacity $\alpha_i^{k-1}$ for obtaining $\alpha_i^k$ as a linear function
of the user satisfaction; formally, $\varepsilon_i^k = \sigma_i^{k-1} - \frac{1}{2}$. This formula shows that
the more $u_i$ is satisfied, the more $\sigma_i^{k-1}$ is higher than $\frac{1}{2}$ and the more $\alpha_i^k$
is increased w.r.t. $\alpha_i^{k-1}$. Clearly, other functions could be adopted for $\varepsilon_i^k$, e.g.,
logarithmic, quadratic, cubic or exponential functions; however, in our opinion, the linear function is the
most suitable one to satisfy the contrasting exigencies mentioned above and to provide the system with the
correct balance between stability and readiness in adapting its behaviour to user satisfaction (see the
Appendix for an experimental analysis of this topic).

\item If, during the execution of $Q_i^{k-1}$, the number of rejected recommendations has been greater than
the number of the accepted ones (i.e., $\sigma_i^{k-1} < \frac{1}{2}$), then it is possible to argue that
$u_i$ has disliked system recommendations, since he refused most of them, and that he is interested to
examine fewer proposals during the next queries. As previously pointed out, the number of system
recommendations depends on $\alpha_i^k$; therefore, in order to make the system to recommend fewer proposals
to $u_i$, it is necessary that $\alpha_i^k$ is lower than $\alpha_i^{k-1}$. For the same reasoning introduced
in the previous case, the most suitable function for defining the decrease $\varepsilon_i^k$ to apply to
$\alpha_i^{k-1}$ for obtaining $\alpha_i^k$ appears the linear one. More formally, the decrease
$\varepsilon_i^k$ can be defined as $\varepsilon_i^k = \frac{1}{2} - \sigma_i^{k-1}$.

\item The last possible situation happens when the number of accepted proposals has been equal to the number
of the rejected ones (i.e., $\sigma_i^{k-1} = \frac{1}{2}$). This situation can be seen as a specific case of
both the first and the second situations described above. If it is considered as a particular case of the
first (resp., second) situation, the increase (resp., decrease) $\varepsilon_i^k$ is equal to 0 and,
consequently, it is possible to conclude that the number of recommendations should be neither increased nor
decreased; this implies that the audacity coefficient should be maintained constant.

\end{itemize}

The previous reasoning leads to the following formula for computing the new value of $\alpha_i^k$:

\begin{center}
$\alpha_i^k = \left\{
\begin{array}{ll}
min\{1, \ \alpha_i^{k-1}+\varepsilon_i^k\} & \mbox{if } \sigma_i^{k-1} > \frac{1}{2} \\
\alpha_i^{k-1} & \mbox{if } \sigma_i^{k-1} = \frac{1}{2}\\
max\{0, \ \alpha_i^{k-1}-\varepsilon_i^k\} & \mbox{if } \sigma_i^{k-1} < \frac{1}{2} \\
\end{array}
\right. $
\end{center}

This formula is valid when $u_i$ exploits the system at least for the second time. During the first query
$\alpha_i^1$ is set equal to a constant value $\overline{\alpha}$. An analysis of the impact of
$\overline{\alpha}$ on the system performance is reported in the Appendix.

\paragraph{Strategy 2} {\em 2-Least Square Error} (2-LSE).

This strategy is based on the Least Square Error technique (LSE) that is largely exploited in several
research areas, such as {\em manufacturing engineering} \cite{GiBaLi02}, {\em bioinformatics}
\cite{SiSaSa03}, {\em econometrics} \cite{Yikang98}, and so on. LSE considers two sets $X =
\{x_1,x_2,\ldots,x_N\}$ and $Y = \{y_1,y_2,\ldots,y_N\}$ and a function  $y = f (x, a_0, a_1, \ldots, a_p)$
depending on $p+1$ {\em unknown} parameters but having a {\em predefined} form (e.g., $f$ might be a
polynomial, a linear combination of exponentials, a sinusoid, and so on). The goal of LSE is to identify the
values $a_0^*$, $a_1^*$, $\ldots$, $a_p^*$ of the parameters $a_0, a_1, \ldots, a_p$ such that the {\em
residual} $R = \sum\limits_{l=1}\limits^{N} \left(y_l-f(x_l,a_0^*,a_1^*,\ldots,a_p^*)\right)^2$ is minimum.
In such a case it is said that $f(x, a_0^*,a_1^*,\ldots, a_p^*)$ is the function that best ``fits'' the sets
$X$ and $Y$, in the least squares sense.

In order to compute $\alpha_i^k$, this strategy:

\begin{itemize}

\item Applies the LSE technique for determining the values $a_0^*, \ldots, a_p^*$ of the parameters $a_0,
\ldots, a_p$ of the fitting function. To this purpose, it exploits the sets $X = \{ \alpha_i^1, \alpha_i^2,
\ldots, \alpha_i^{k-1} \}$ and $Y = \{ \sigma_i^1, \sigma_i^2, \ldots, \sigma_i^{k-1} \}$. Observe that, as a
consequence of this choice for the sets $X$ and $Y$, the fitting function represents the user satisfaction
against the system audacity.

\item Sets $\alpha_i^k$ to the value of the audacity corresponding to the maximum value\footnote{Observe
that, since the independent variable (i.e., the audacity) belongs to the compact interval $[0,1]$, the
fitting function has always a maximum within this interval.} of the fitting function and, consequently, to
the maximum value of the user satisfaction.

\end{itemize}

A crucial problem in the implementation of the LSE strategy is the choice of both the shape and the
parameters of the fitting function. Such a choice must minimize the residual $R$; as a consequence, it
requires to compute the partial derivatives of $R$ and to solve the set of equations
$\frac{\partial{R}}{\partial{a_j}} = 0, \forall j = 0,\ldots,p$.

If we use, as fitting function, a combination of sinusoids, logarithms and exponentials, these equations
generally define a non-linear system that, often, cannot be exactly solved; in this case, even the
calculation of an approximate solution requires an heavy computational effort.

On the contrary, if the fitting function has a polynomial shape, i.e., $f(x, a_0, a_1, \ldots, a_p) = a_0
x^p+ a_1 x^{p-1} + \ldots + a_p$, it is possible to show that these equations are equivalent to the {\em
linear system} \cite{GoVa96} {\bf Ax = b}, where {\bf A} is a matrix called {\em Vandermonde matrix}. It is
possible to show that at least an approximate solution of this system can be computed in a polynomial time;
generally, the exact solution can be computed in $O(p^2)$ steps \cite{GoVa96}; as a consequence, a polynomial
shape for the fitting function appears well suited for our purposes.

The next step of our activity consists of determining the order of the polynomial to be used as fitting
function. To this purpose observe that LSE must compute the audacity value that maximizes the fitting
function $f(x)$; to compute this value we need to determine the derivative $f'(x)$ of $f(x)$ and to solve the
equation $f'(x)=0$. Now, if $f(x)$ is a $p^{th}$ order polynomial, $f'(x)$ will be a $(p-1)^{th}$ order
polynomial; as a consequence, we have that: {\em (i)} if $p = 2$ (i.e., if the fitting function is a
parabola), it is possible to directly obtain an exact solution of $f'(x)=0$ by solving a first order
equation; {\em (ii)} if $p = 3$ (i.e., if the fitting function is a cubic), it is possible to obtain an exact
solution of $f'(x)=0$ by solving a second order equation; {\em (iii)} if $p > 3$, the computation of the
roots of $f'(x)$ can be performed by means of numerical (and approximate) techniques; with regard to this, it
was shown that the time necessary for finding all the roots of $f'(x) = 0$ with a maximum error equal to
$\frac{1}{2^b}$ is $O(p \log p^4 + p \log^2 p \log b)$ \cite{Pan97}.

This analysis shows that, for obtaining a correct (i.e., not approximate) solution of the equation $f'(x)=0$,
the fitting function must be either a parabola or a cubic. Since the accuracy ensured by the polynomial of
degree 2 (i.e., the parabola) is high \cite{Bertsekas99}, and since the computation time it requires for
solving $f'(x)=0$ is lower than that necessary if the fitting function is a cubic, we have adopted the
parabola as fitting function; in this case the LSE technique is called 2-LSE technique.

Therefore, in our case, the fitting function is $y=a_0x^2+a_1x+a_2$; this choice implies that the 2-LSE
strategy is valid only when $u_i$ exploits the system at least for the fourth time. During the first three
queries $\alpha_i^1, \alpha_i^2, \alpha_i^3$ are set equal to three constant values $\overline{\alpha'},
\overline{\alpha''}, \overline{\alpha'''}$, respectively. An analysis of the impact of $\overline{\alpha'},
\overline{\alpha''}, \overline{\alpha'''}$ on the system performance is reported in the Appendix.

\paragraph{Strategy 3} {\em Weighted Sum} (WS).

In order to compute $\alpha_i^k$, this strategy applies the PNF and the 2-LSE strategies and obtains two
audacity values $\alpha_i^{k,PNF}$ and $\alpha_i^{k,2-LSE}$. The overall audacity value $\alpha_i^k$ is
obtained by computing a suitable weighted mean of $\alpha_i^{k,PNF}$ and $\alpha_i^{k,2-LSE}$; specifically,
$\alpha_i^k = \gamma \times \alpha_i^ {k,PNF} + (1-\gamma) \times \alpha_i^{k,2-LSE}$. Here, $\gamma$ is a
coefficient ranging in the real interval $[0,1]$. Specifically, if $\gamma = 1$, the WS strategy coincides
with the PNF strategy; on the contrary, if $\gamma = 0$, the WS strategy coincides with the 2-LSE strategy.
In our experiments we have considered different values for $\gamma$ and we have studied their impact on the
system performance (see the Appendix for more details); in addition, we have examined how to dynamically
compute $\gamma$ for maximizing the system performances.

\section{Experimental Results}
\label{sec:Experimental}

In this section we describe in detail the various experiments we have conducted for testing the performance
of our system. Specifically, in Section \ref{sub:characteristics} we describe the characteristics of both the
users and the job proposals considered in our tests; Section \ref{sub:evmetrics} presents the accuracy
measures we have chosen for testing our approach. Section \ref{sub:prerecdiffdomain} is devoted to make a
comparison of the three strategies for the audacity computation. In Section \ref{sub:prerecspecialized} we
evaluate the performance of our system in different application domains. Finally, Section
\ref{sub:comparison} is devoted to present an experimental comparison of our system with other e-recruitment
ones.

In order to perform our tests, we have designed a prototype implementing our approach. This prototype has
been developed in JADE\footnote{JADE is an open source platform; it can be downloaded from {\tt
http://sharon.cselt.it/projects/jade.}} (Java Agent DEvelopment framework), a software framework conceived
for supporting the implementation of agent-based applications in compliance with FIPA specifications
\cite{FIPA02}.

\subsection{Characteristics of the users and job proposals}
\label{sub:characteristics}

In our tests we have considered a set of users $USet = \{u_1, u_2, \ldots, u_{200}\}$ consisting of 200
volunteers. As for their distribution against their age we have that: {\em (i)} 21.00\% of them was 18-24
years old; {\em (ii)} 39.00\% of them was 25-34 years old; {\em (iii)} 25.50\% of them was 35-44 years old;
{\em (iv)} 13.50\% of them was 45-54 years old; {\em (v)} 1.00 \% of them was more than 54 years old.

As for their distribution against their past usage to commercial e-recruitment systems we have that: {\em
(i)} 11.00\% of them had never used a commercial e-recruitment system; {\em (ii)} 10.00\% of them exploited
it very rarely; {\em (iii)} 10.00\% of them used it rarely; {\em (iv)} 25.50\% of them exploited it
sometimes; {\em (iv)} 31.00\% of them exploited it often; {\em (v)} 12.50\% of them used it very often.

The available job proposals, extracted from the sites {\small \tt http://www.jobpilot.co.uk} and {\small \tt
http://www.careerbuilder.com} were about 3000. They belonged to different domains; specifically, reference
domains where: {\em (i)} ``Information Technology'' for 19.29\% of them; {\em (ii)} ``Health Care
Management'' for 17.02\% of them; {\em (iii)} ``Finance'' for 15.84\% of them; {\em (iv)} ``Engineering'' for
12.07\% of them; {\em (v)} ``Banking'' for 10.04\% of them; {\em (vi)} ``Legal'' for 7.02\% of them; {\em
(vii)} ``Real Estate Management'' for 5.26\% of them; {\em (viii)} ``Others'' for 13.46\% of them.

\subsection{Evaluation Measures} \label{sub:evmetrics}

In our tests we have computed two widely accepted evaluation measures, namely {\em Precision} and {\em
Recall}. Precision is defined as the share of job proposals accepted by $u_i$ among those recommended by the
system; Recall is the share of job proposals suggested by the system among those of interest for $u_i$.

These parameters have been computed as follows:

\begin{itemize}

\item  Each user $u_i \in USet$ was asked to submit a query $Q_i^k$ and the corresponding $JPTempList_i^k$
was obtained (see Section \ref{sec:RecommenderAgent}).

\item $u_i$ was asked to identify the subset $UserList_i^k \subseteq JPTempList_i^k$ of job proposals that he
considered interesting.

\item Our prototype was run for obtaining the set $JPList_i^k \subseteq JPTempList_i^k$ of job proposals to
be recommended to $u_i$.

\item The Precision $Pre_i^k$ and the Recall $Rec_i^k$ relative to the $k^{th}$ query have been obtained by
applying the formulas: $Pre_i^k = \frac{|JPList_i^k \cap UserList_i^k|}{|JPList_i^k|}$, $ Rec_i^k =
\frac{|JPList_i^k \cap UserList_i^k|}{|UserList_i^k|}$.

\item Finally, the {\em Average Precision} $AvgPre^k$ and the {\em Average Recall} $AvgRec^k$ relative to the
$k^{th}$ query have been obtained as $ AvgPre^k = \frac{\sum_{i=1..|USet|} Pre_i^k}{|USet|}$, $AvgRec^k =
\frac {\sum_{i=1..|USet|} Rec_i^k}{|USet|}$.

\end{itemize}

Observe that Precision, Recall, Average Precision and Average Recall belong to the real interval $[0,1]$;
specifically, the higher these coefficients are the better the system works.

\subsection{Experimental comparison of the three strategies implemented in our approach}
\label{sub:prerecdiffdomain}

In this section we experimentally compare the strategies for the audacity computation described in Section
\ref{sec:personalization}. Actually, it is necessary to examine only the PNF and the 2-LSE strategies; in
fact, the WS strategy is a weighted mean of the two other ones. In the test of the PNF strategy we have set
$\overline{\alpha}=0.55$ whereas in the test relative to the 2-LSE strategy we have set $\overline{\alpha'} =
0.5, \overline{\alpha''}=0.6, \overline{\alpha'''}=0.4$ (see the Appendix).

\begin{figure}[t]
      \centerline{\psfig{figure=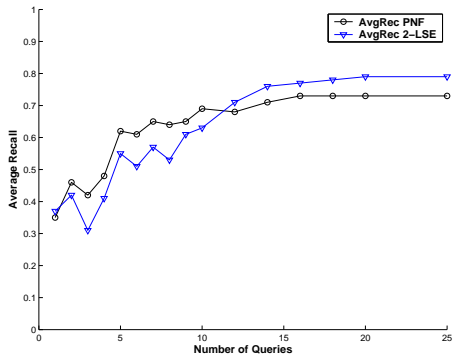,width=6cm}}
      \caption{Average Recall for the PNF and the 2-LSE strategies}
      \label{fig:avgrec}
\end{figure}

For each strategy we have computed the Average Precision and the Average Recall against the number of
queries. The obtained results are shown in Figures \ref{fig:avgpre} and \ref{fig:avgrec}. From the analysis
of these figures we can conclude that:

\begin{figure}[t]
      \centerline{\psfig{figure=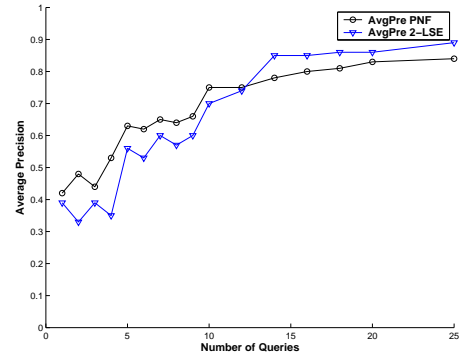,width=6cm}}
      \caption{Average Precision for the PNF and the 2-LSE strategies}
      \label{fig:avgpre}
\end{figure}

\begin{itemize}

\item If the number of queries carried out by users is low (i.e. lesser than 10), the PNF strategy shows a
better performance than the 2-LSE strategy; in fact, its Average Precision and its Average Recall increase up
to 0.75 and 0.69, respectively. This behaviour can be motivated by considering that, before 10 queries have
been performed, the 2-LSE strategy has still not completed the audacity ``tuning'' and, hence, the
corresponding results are not particularly satisfactory.

\item If the number of queries carried out by users is high (i.e. greater than 15), the performance
improvements of the PNF strategy are small. On the contrary, the 2-LSE strategy shows a very good
performance; as an example, after 15 queries, the Average Precision and the Average Recall are 0.85 and 0.76,
respectively. This behaviour can be motivated by considering that the 2-LSE strategy ``tunes'' the audacity
coefficient on the basis of the user feedback during all the $k-1$ previous queries, whereas the PNF strategy
considers only the last one. As a consequence, a high number of queries allows the 2-LSE strategy to more
precisely predict user expectations.

\item During the initial queries, both Average Precision and Average Recall have an oscillatory behaviour;
this depends on the fact that, initially, both user profiles and user feedbacks are relatively poor.

\item  Finally, we observe that the Average Precision is generally greater than the Average Recall; this
confirms the results obtained in \cite{BrSm03}. As pointed out in this paper, such a feature is to be
considered a positive characteristic for a recruitment system; in fact, in this application context,
Precision is generally assumed to be more important than Recall since the user could be frustrated by many
irrelevant proposals.

\end{itemize}

Finally, it is worth pointing out that the results of this experiment confirm the results of the experiments
about the tuning of $\gamma$ described in the Appendix. In fact, in that case, we have obtained that, for a
small number of queries (i.e., lesser than 10), the value of $\gamma$ should be high (i.e., the WS strategy
should tend to coincide with the PNF strategy); on the contrary, when the number of queries is high, the
value of $\gamma$ should be low (i.e., the WS strategy should tend to coincide with the 2-LSE strategy).

\subsection{Performance of our system in different application domains}
\label{sub:prerecspecialized}

This series of experiments was devoted to determine the performance of our system in several application
domains; some of them were very generic, other ones were more specialized.

\begin{figure*}
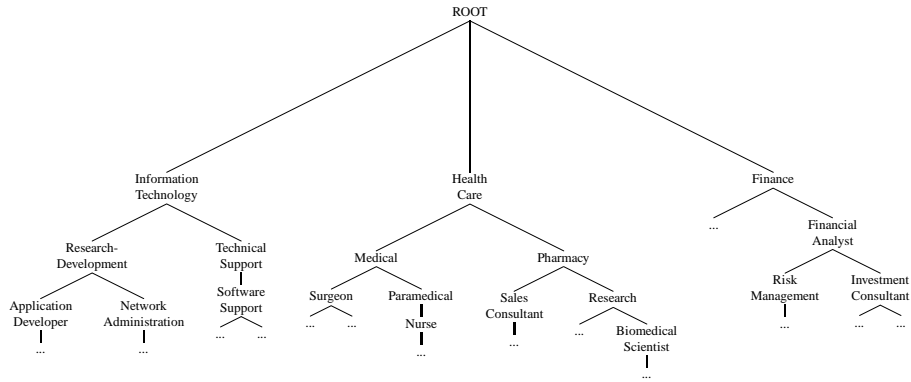

\begin{center}
{\tiny \hskip -0.5in\Tree [.ROOT [.{Information\\Technology} [.{Research-\\Development}
[.{Application\\Developer} {...} ] [.{Network\\Administration} {...} ] ] [.{Technical\\Support}
[.{Software\\Support} {...} {...} ]
  ] ] [.{Health\\Care} [.Medical [.Surgeon {...} {...} ] [.Paramedical
[.Nurse {...} ] ] ] [.Pharmacy [.{Sales\\Consultant} {...} ] [.{Research} {...} [.{Biomedical\\Scientist}
{...} ] ] ] ] [.{Finance} {...} [.{Financial\\Analyst} [.{Risk\\Management} {...} ]
[.{Investment\\Consultant} {...} {...} ] ] ]]}
\end{center}
  \caption{A fragment of a tree classifying job proposals}
  \label{fig:tree of job proposal}
\end{figure*}

In all existing e-recruitment systems, job proposals can be hierarchically organized on the basis of the
application domains they refer to; specifically, the most generic domains (e.g., ``Health-Care'' or
``Information Technology'') are located on the top of the hierarchy; on the contrary, the most specialized
ones (like ``Biochemical Scientist'') are located at the bottom. This hierarchy can be graphically
represented by means of a tree; each node of the tree, with the exception of the root, represents a domain; a
fragment of this tree is graphically shown in Figure \ref{fig:tree of job proposal}. With regard to this
classification tree, we say that the {\em specialization level} of a domain $D$ is $j$ if the depth of the
node associated with $D$ is $j$. As an example, the specialization level of the domain ``Application
Developer'' is 3, since the depth of the node representing this domain is 3.

In this test we have computed the Average Precision and the Average Recall obtained by our system when
applied to four domains, namely ``Information Technology'', ``Pharmacy'', ``Software Support'' and
``Biomedical Scientist'', characterized by different specialization levels. During this computation we have
used the WS strategy by setting $\gamma(k)=max \left\{0,1-\left(\frac{k-1}{25}\right) \right\}$,
$\overline{\alpha} =0.55$ and $\overline{\alpha'} =0.5, \overline{\alpha''} = 0.6, \overline{\alpha'''}=0.4$.
The obtained results are shown in Table \ref{tab:PerformanceDomain}.

\begin{table*}[t]
\begin{center}
  \caption{Performance of our system in different application domains}
  \label{tab:PerformanceDomain}
{\tiny
\begin{tabular}{||c|c|c|c|c|c|c|c|c|c|c|c||}
\hline \hline {\em Domain} & {\em Specialization Level} & \multicolumn{2}{|c}{\em Query 5} & \multicolumn{2}{|c}{\em Query 10} & \multicolumn{2}{|c}{\em Query 15} &\multicolumn{2}{|c}{\em Query 20} & \multicolumn{2}{|c||}{\em Query 25} \\
     \cline{3-12}
          && {\em Avg Pre} & {\em Avg Rec} & {\em Avg Pre} & {\em Avg Rec} & {\em Avg Pre} & {\em Avg Rec} & {\em Avg Pre} & {\em Avg Rec} & {\em Avg Pre} & {\em Avg Rec} \\
\hline {\em Information Technology} & 1  & 0.73 & 0.67   & 0.76 & 0.70   & 0.74 & 0.68   & 0.72 & 0.65   & 0.68 & 0.61 \\
\hline {\em Pharmacy} & 2                & 0.68 & 0.63   & 0.72 & 0.69   & 0.79 & 0.71   & 0.78 & 0.70   & 0.77 & 0.69 \\
\hline {\em Software Support} & 3        & 0.54 & 0.52   & 0.61 & 0.58   & 0.80 & 0.72   & 0.85 & 0.73   & 0.86 & 0.75 \\
\hline {\em Biomedical Scientist} & 4    & 0.49 & 0.46   & 0.60 & 0.56   & 0.81 & 0.73   & 0.86 & 0.76   & 0.90 & 0.82 \\
\hline \hline
\end{tabular}
}
\end{center}
\end{table*}

From the analysis of this table it is possible to observe that, in general domains (i.e., domains at a
specialization level 1 or 2), our system rapidly (i.e., after a few number of queries) obtains good results;
however, this performance slightly worsens when the number of queries increases. On the contrary, in specific
domains, our system needs several queries for achieving good results but, after this initial phase, it
maintains, and even improves, its performance.

This behaviour can be motivated by the following reasoning: during the first queries, users interested in
general domains do not have a precise idea about their needs and, consequently, many of the provided
recommendations appear interesting to them. On the contrary, users interested in specific domains have a
precise idea of their desires already during the initial phase; as a consequence, it is more difficult to
satisfy them during the first queries. When the number of queries increases, users of general domains
``ripen'' a more precise idea of their needs, but the application domain they are interested in is too
generic to precisely satisfy their requirements; however, Table~\ref{tab:PerformanceDomain} shows that, even
in these conditions, our system achieves quite good results. On the contrary, after many queries, the
profiles of the users of specific domains are quite rich and this allows our system to precisely identify
their needs and to significantly reduce the job proposals search space.

\subsection{Experimental comparison of our system with other e-recruitment prototypes}
\label{sub:comparison}

We have experimentally compared our system with other, already available, ones. Specifically, the systems
that we have taken into account were: {\em (i) CareerBuilder} - {\tt \small www.careerbuilder.com}, {\em (ii)
1job} - {\tt \small www.1job.co.uk}, {\em (iii) Job Search} - {\tt \small www.jobsearch.co.uk}, {\em (iv)
JobPilot} - {\tt \small www.jobpilot.co.uk}, and {\em (v) Fish4Jobs} - {\tt \small www.fish4.co.uk/iad/jobs}.
These systems share various similarities; specifically, they operate on the Internet and manage an internal,
relational database storing all available job proposals. In addition, they provide job seekers with a
graphical interface to retrieve the job proposals they are interested in; a job seeker can use this interface
to specify some constraints (e.g., the place where he would like to work), as well as a list of keywords
describing the job typology of his interest. Finally, some of these systems allow each user to construct a
simple profile by filling a questionnaire; each profile stores user skills (e.g., the foreign languages he
knows) as well as a brief curriculum vitae. Since the queries submitted by a job seeker consist of a list of
keywords, each system must transform them into SQL queries; for this reason, the underlying DBMS must
incorporate Information Retrieval (hereafter, IR) algorithms \cite{HrGrPa03}.

Our activity aimed at comparing the accuracy of our system against that of the other systems mentioned
previously. To this purpose we asked each user of $USet$ to submit some queries; these were processed by the
systems into examination and the corresponding Average Precision and Average Recall were computed. As for our
system we have chosen the WS strategy and we have set $\gamma(k)=max \left\{0,1-\left(\frac{k-1}{25}\right)
\right\}$, $\overline{\alpha} =0.55$ and $\overline{\alpha'} =0.5, \overline{\alpha''} = 0.6,
\overline{\alpha'''}=0.4$. The results we have obtained are shown in Table~\ref{tab:PerformanceComparison}.
From the analysis of this table, we can observe that the accuracy of our system is quite satisfying. This
interesting result can be justified by the following reasoning:

\begin{table*}[t]
\begin{center}
 \caption{Average Precision and Average Recall of the systems into examination}
\label{tab:PerformanceComparison} {\tiny
\begin{tabular}{||c|c|c|c|c|c|c|c|c|c|c||}
\hline \hline {\em System} & \multicolumn{2}{|c}{\em Query 5} & \multicolumn{2}{|c}{\em Query 10} &
\multicolumn{2}{|c}{\em Query 15}
&\multicolumn{2}{|c}{\em Query 20} & \multicolumn{2}{|c||}{\em Query 25}\\
\cline{2-11} & {\em Avg Pre} & {\em Avg Rec} & {\em Avg Pre} & {\em Avg Rec}
& {\em Avg Pre} & {\em Avg Rec} & {\em Avg Pre} & {\em Avg Rec} & {\em Avg Pre} & {\em Avg Rec}\\
\hline {\em Our System}     &0.62   &0.61   &0.73   &0.66   &0.81   &0.74   &0.86   &0.76   &0.87   &0.79\\
\hline {\em CareerBuilder}  &0.50   &0.48   &0.61   &0.52   &0.68   &0.57   &0.72   &0.63   &0.73   &0.65\\
\hline {\em 1job}           &0.52   &0.46   &0.63   &0.57   &0.71   &0.66   &0.68   &0.67   &0.70   &0.63\\
\hline {\em Job Search}     &0.54   &0.44   &0.64   &0.60   &0.69   &0.64   &0.70   &0.65   &0.73   &0.68\\
\hline {\em JobPilot}       &0.44   &0.51   &0.47   &0.53   &0.59   &0.54   &0.67   &0.60   &0.71   &0.66\\
\hline {\em Fish4Jobs}      &0.47   &0.48   &0.50   &0.52   &0.60   &0.55   &0.63   &0.62   &0.69   &0.63\\
\hline \hline
\end{tabular}
}
\end{center}
\end{table*}

\begin{itemize}

\item In order to score available job proposals, the ranking function adopted by our system considers several
aspects of the profile of a user, as well as his reaction to past proposals (see
Sections~\ref{subsec:uabehavior} and \ref{sub:RABehaviour}); on the contrary, the ranking functions adopted
by the other systems into examination consider only some constraints specified by the user (e.g., the salary
he would like to earn). Taking into account a large number of issues makes our system more sensible to the
real user exigencies; this justifies the improvement of the Average Precision w.r.t. the other systems into
examination.

\item Our system suggests to users not only the job proposals exactly matching their queries but also those
ones someway semantically related to them. Hence, it is capable of suggesting potentially interesting job
proposals that cannot be revealed by traditional IR algorithms; this implies an improvement of the Average
Recall w.r.t. the other systems into examination.

\item Our system is capable of considering job proposals available in several other systems; this contributes
to improve both the Average Precision and the Average Recall w.r.t. the other systems into examination.

\end{itemize}

However, neither Precision nor Recall alone are capable of completely capturing the ``quality'' of the
proposals provided by e-recruitment systems. As an example, a key aspect of these systems regards their
capability of correctly scoring their job proposals. To better clarify this concept, consider two
e-recruitment systems, $ERS_1$ and $ERS_2$, that receive the same query from a user $U_i$ and return the same
set of results but in a {\em reverse order}: in other words, $ERS_1$ returns the list $\langle JP_1, JP_2,
\ldots, JP_{n-1}, JP_n \rangle$ whereas $ERS_2$ returns the list $\langle JP_n, JP_{n-1}, \ldots,JP_2, JP_1
\rangle$; this implies that $ERS_1$ (resp., $ERS_2$) rates $JP_1$ (resp., $JP_n$) as the most relevant
proposal for $U_i$ and $JP_n$ (resp., $JP_1$) as the least relevant one. Finally, assume that $U_i$ fully
agrees with the suggestions provided by $ERS_1$. In this case $ERS_1$ and $ERS_2$ return the same proposals
and, consequently, obtain the same Average Precision and the same Average Recall; however, from the previous
reasoning, it is possible to conclude that the ``quality'' of suggestions provided by $ERS_2$ is lower than
the quality of suggestions provided by $ERS_1$: in fact, $U_i$ must browse the whole list of suggestions
provided by $ERS_2$ to find the job proposal of his maximum interest and this might be both a boring and a
time-consuming activity.

To measure the {\em ranking capability} of an e-recruitment system we adopted the {\em Newell Distance}
\cite{Newell97}, a parameter defined in User Modelling theory. To define the Newell Distance associated with
a user $U_i$ and a system to evaluate $S_j$ we need to consider a set $JPSet$ of test job proposals and a
pair of functions, $sys$ and $usr$; $sys$ (resp., $usr$) takes a job proposal $JP_k \in JPSet$ as input and
returns an integer $p$ belonging to the set $\{1, 2, \ldots, |JPSet| \}$ as output; $p$ indicates that,
according to $S_j$ (resp., $U_i$), $JP_k$ has the $p^{th}$ highest score among all job proposals existing in
$JPSet$.

The Newell Distance ${\cal N}_{ij}$ associated with $U_i$ and $S_j$ is defined as {\small ${\cal N}_{ij} =
\sum\limits_{JP_k \in JPSet} \mid w(usr(JP_k)) \times usr(JP_k) - w(sys(JP_k)) \times sys(JP_k) \mid $}. Here
$w(i)$ is a weighting function defined as: $w(i) = \left(\frac{\mid JPSet \mid - i}{i}\right)^2$; it assumes
its maximum value when $i =1$ and decreases when $i$ increases; this implies that it gives a different
importance to the errors possibly made by the system to evaluate; specifically, if a system is unable to
detect the most important job proposal, it commits a more serious error than if it is not capable of
identifying the least relevant ones.

The Newell Distance can be properly normalized in such a way to lie in the real interval [0,1]; specifically,
if we indicate by ${\cal N}^{max}$ the maximum value of the Newell Distance measured over all users and
systems into consideration, the normalized Newell Distance can be defined as $ \hat{{\cal N}_{ij}} =
\frac{{\cal N}_{ij}}{{\cal N}^{max}}$. Observe that the lesser $\hat{{\cal N}_{ij}}$ is, the better a system
works.

In Table~\ref{tab:Newell} we report the normalized Newell Distance, averaged on all users, obtained for the
systems into examination. From the analysis of this table we observe that the Newell Distance obtained by our
system is smaller than that obtained by the other systems into evaluation. In our opinion, this result is
motivated by considering that:

\begin{table}[t]
\begin{center}
\caption{Average normalized Newell distance achieved by the systems into examination} \label{tab:Newell}
{\tiny
\begin{tabular}{||c|c|c|c|c|c||}
\hline \hline {\em System} & {\em Query 5} & {\em Query 10} &{\em Query 15} & {\em Query 20} & {\em
Query 25}\\
\hline {\em Our System}     &0.27   &0.20   &0.13   &0.09   &0.06\\
\hline {\em CareerBuilder}  &0.40   &0.31   &0.19   &0.15   &0.12\\
\hline {\em 1job}           &0.38   &0.36   &0.22   &0.14   &0.11\\
\hline {\em Job Search}     &0.33   &0.29   &0.21   &0.14   &0.09\\
\hline {\em JobPilot}       &0.39   &0.33   &0.29   &0.24   &0.15\\
\hline {\em Fish4Jobs}      &0.38   &0.31   &0.25   &0.19   &0.13\\
\hline \hline
\end{tabular}
}
\end{center}
\end{table}

\begin{itemize}

\item The parameters adopted by our system for scoring available job proposals are strictly related to the
profile of a user, as well as to his reaction to past proposals; this enhances its capability of correctly
identifying the most relevant job proposals.

\item Our system continuously monitors user behaviour and is capable of detecting if the relevance of a job
proposal for a given user changes over time. On the contrary, the other systems we have considered in our
experiments do not consider the possible modifications of user desires over time.

\end{itemize}

From the previous experiments we can conclude that the recommendations provided by our system are more
accurate than those supplied by the other systems into examination; this improvement is mainly due to quite a
sophisticated management of information about user profiles and past behaviour. This information is combined
with classical IR techniques exploited also by the other systems into consideration and allows the most
appropriate answers for user needs to be found.

\begin{figure}[t]
\centerline{\psfig{figure=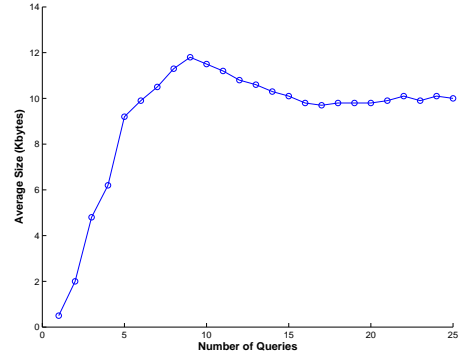,width=6cm}} \caption{Average size of user profiles against the number
of queries carried out by users} \label{fig:avgusrprofile}
\end{figure}

On the other hand, our system requires an additional amount of space for storing user profiles; this is a
problem affecting all systems handling and exploiting user profiles. In order to quantify the space
requirements of our system for storing user profiles we have performed a final experiment. Specifically, we
have computed the average size of user profiles (expressed in Kbytes) against the number of queries carried
out by users. The obtained results are shown in Figure \ref{fig:avgusrprofile}.

From the analysis of this figure we can observe that, initially, user profiles are generally ``poor'' and,
consequently, the storage space they require is negligible; when a user poses his queries, the system
enriches the corresponding profile by inserting new topics. After a ``reasonably large'' number of queries
(i.e., after about 10 queries) user profiles become rich and occupy a certain amount of space; however, this
space occupation does not increase during the next queries; in fact, when user profiles become excessively
large, the system activates a pruning task in such a way that the number of new topics inserted in a profile
is approximately equal to the number of topics removed from it; as a consequence, after 15 queries, the
average size of user profiles remains quite constant (specifically, it is about 10 Kbytes).

The previous analysis shows that the quantity of space our system needs for storing user profiles is limited
and acceptable; moreover, since user profiles are implemented in XML, it is possible to apply the
methodologies illustrated in \cite{Jagadish*02} for handling their space occupation in a more efficient way.

\section{Related Work}
\label{sec:rw}

Even if e-recruitment is quite a novel research area, several systems devoted to handle such an activity have
been already presented in the literature. In this section we aim at positioning our system amongst other
related ones. In order to carry out this activity, we have considered three terms of comparison, namely: {\em
(i) Purpose}: e-recruitment systems can be classified as company oriented, if they aim at supporting
companies in the selection of new candidates, and seeker oriented, if they support users looking for new job
proposals; {\em (ii) Architecture}: e-recruitment systems could be centralized, if a single computational
entity is in charge of managing all activities, distributed, if tasks are partitioned among several
computational entities, and mixed, if some tasks are performed in a centralized fashion and other ones are
carried out in a distributed way; {\em (iii) User Profile Construction}: e-recruitment systems might be
obtrusive, if they interact with the user for constructing his profile, or unobtrusive, if they learn the
profile of a user by monitoring his accesses to the system.

On the basis of these terms of comparison our system can be considered {\em seeker oriented}, {\em mixed} and
{\em unobtrusive}.

In the following we compare our approach with other related ones and highlight similarities and differences
existing among them. Table \ref{tab:ComparisonRecruitment} provides a summary of the in-depth comparison we
carry out in the following subsections.

\begin{table*}[t]
\begin{center}
\caption{A comparison of some e-recruitment systems} \label{tab:ComparisonRecruitment} {\tiny
\begin{tabular}{||c||c|c|c|c||}

\hline \hline
{\em System}         & {\em Purpose} & {\em Architecture}& {\em User Profile construction} & {\em Additional Features}  \\
\hline\hline
{\em Our approach} & Seeker Oriented & Mixed        & Unobtrusive & It exploits Agent Technology and XML  \\
\hline {\em Supjarerndee et al. } \cite{SuTePh02}& Seeker Oriented & Centralized & Obtrusive &  It considers the personal traits of a user\\
\hline {\em Farber et al.} \cite{FaKeWe03} & Seeker Oriented & Centralized & Unobtrusive & It is an hybrid recommender system\\
\hline {\em CASPER} \cite{BrSm03,RaBrSm00}& Seeker Oriented & Mixed & Unobtrusive & It separately exploits content based and collaborative filtering techniques\\
\hline {\em  Personnel Mall} \cite{GaNi98} & Company Oriented &Distributed & Obtrusive & It exploits the agent technology and a matching algorithm\\
\hline {\em Dafoulas et al.} \cite{DaNiTu03} & Company Oriented & Centralized & Obtrusive & It allows online interviews to be performed\\
\hline {\em CommOnCV} \cite{HaLeTr02}& Seeker Oriented & Centralized & Obtrusive & It exploits Semantic Web languages to model a curriculum vitae\\
\hline {\em e-SRS} \cite{KhGo03} & Company Oriented & Distributed & Obtrusive & It exploits both psychometric techniques and clustering algorithms to recruit candidates\\
\hline\hline

\end{tabular}
}
\end{center}
\end{table*}

\subsection{Approach of \cite{SuTePh02}}

\cite{SuTePh02} proposes a recommender system devoted to support users looking for new job proposals. The
system operates as follows: {\em (i)} it classifies each user according to his personal traits (e.g.,
shy/sociable or talktive/taciturn); {\em (ii)} a user queries it to retrieve new job proposals; {\em (iii)}
it ranks retrieved job proposals according to their similarity to the personal traits of the user. There are
some similarities between our approach and \cite{SuTePh02}; in fact, both of them: {\em (i)} are seeker
oriented; {\em (ii)} can cooperate with existing e-recruitment Web sites; {\em (iii)} classify job proposals
according to their relevance for the user. As for differences, we may notice that: {\em (i)} in
\cite{SuTePh02} user profiles are constructed by means of psychological tools (e.g., questionnaires) whereas
our system obtains information about users unobtrusively, by watching their interactions with it; {\em (ii)}
in \cite{SuTePh02} no mechanisms for user profile update are provided; {\em (iii)} the architecture of
\cite{SuTePh02} is centralized whereas our system is mixed, since the Recruitment Agent performs most of the
activities connected with the selection of the job proposals best fitting user exigencies, but User Agents,
Wrapper Agents and Company Agents continuously cooperate with the Recruitment Agent for performing the whole
recruitment task (see Section \ref{sec:Overview}). Interestingly enough, our system might inherit some
features typical of the approach of \cite{SuTePh02}; specifically, the user profiles managed in our system
might include information about user personal traits, analogous to that handled in \cite{SuTePh02}, to
enhance the job search process.

\subsection{Approach of \cite{FaKeWe03}}

In \cite{FaKeWe03} the authors present a probabilistic e-recruitment system that uses both content-based and
collaborative filtering techniques to produce recommendations. There are some similarities between our
approach and that described in \cite{FaKeWe03}; more specifically, both of them: {\em (i)} provide a
mechanism for automatically and unobtrusively learning user preferences; {\em (ii)} take into account the
user feedback that influences the system behaviour: specifically, in our approach, the knowledge of the job
proposals accepted/rejected by the user is exploited for automatically tuning the system audacity;
analogously, in \cite{FaKeWe03}, user preferences influence the coefficients of the probabilistic model
adopted for deriving recommendations; {\em (iii)} are seeker oriented. As for differences, we may notice
that: {\em (i)} \cite{FaKeWe03} exploits a probabilistic model for deducing new recommendations; such an
approach is very refined but it might be excessively time-consuming; {\em (ii)} our system is mixed whereas
the approach of \cite{FaKeWe03} is centralized; {\em (iii)} the approach of \cite{FaKeWe03} is hybrid,
whereas our own is content-based.

\subsection{CASPER}

In \cite{BrSm03,RaBrSm00} the system CASPER is described. CASPER is a client-server system that {\em
separately} exploits {\em collaborative filtering} and {\em content-based} techniques to provide new job
recommendations. There are some similarities between our approach and CASPER; in fact, both of them:  {\em
(i)} are seeker oriented; {\em (ii)} construct user profiles by unobtrusively monitoring user activities;
{\em (iii)} follow similar criteria for estimating user preferences; {\em (iv)} are mixed. As for differences
we may notice that: {\em (i)} CASPER can operate either as a content-based or as a collaborative filtering
recommender system; on the contrary, our system is content-based; {\em (ii)} CASPER does not provide tools
for modelling the semantics of a job proposal; on the contrary, our approach uses an XML Schema for
representing the main features of each job proposal; these features play a key role during the recommendation
process.

\subsection{Personnel Mall}

In \cite{GaNi98} the {\em Personnel Mall} system is proposed. {\em Personnel Mall} is an agent based platform
conceived for matching job seekers with companies in a distributed (and electronic) job marketplace.
Specifically, {\em Personnel Mall} defines a set of rules to represent and manage the preferences/exigencies
of job seekers and companies and uses them during the matching process. There are some similarities between
our approach and {\em Personnel Mall}; in fact, both of them: {\em (i)} use the agent technology to perform
e-recruitment activities; {\em (ii)} model job marketplace as a complex system in which heterogeneous
components (i.e., companies and individuals) incessantly interact. As for differences, we may observe that:
{\em (i)} {\em Personnel Mall} is company oriented; {\em (ii)} in {\em Personnel Mall} the matching between a
job seeker and a job proposal depends not only on {\em subjective variables} (e.g., the personal interests of
a user) but also on {\em economic parameters} (e.g., the quantity of labor that companies would like to hire
for a certain wage and the quantity of labor that a job seeker would like to supply); {\em (iii)} {\em
Personnel Mall} aims at maintaining the conditions of equilibrium in the job marketplace: as an example, it
is capable of dynamically raising/lowering wages in response to labor surpluses/shortages; such a feature is
not present in our system; {\em (iv) Personnel Mall} does not provide mechanisms for updating the preferences
and the exigencies of job seekers; {\em (v)} {\em Personnel Mall} is obtrusive; {\em (vi)} {\em Personnel
Mall} is distributed; on the contrary, our system is mixed.

\subsection{Approach of \cite{DaNiTu03}}

In \cite{DaNiTu03} an agent based approach for supporting e-recruitment activities is proposed. This system
embodies the functionalities typical of a traditional e-recruitment Web site (e.g., a user can insert/update
his profile, a company can post new job proposals, and so on); moreover, it allows companies to perform an
{\em on-line interview} of candidates; this interview is useful for both companies (that can perform a
preliminary screening of candidates) and job seekers (that can enrich their profiles by storing their past
interviews). It is possible to identify only few similarities between our approach and \cite{DaNiTu03};
specifically, both of them: {\em (i)} define and manage user profiles; {\em (ii)} use the agent technology.
As for differences we observe that: {\em (i)} \cite{DaNiTu03} is company oriented; {\em (ii)} \cite{DaNiTu03}
provides a {\em layered} and {\em centralized} architecture whereas our approach is mixed; {\em (iii)}
\cite{DaNiTu03} is obtrusive.

\subsection{CommOnCV}

In \cite{HaLeTr02} the authors propose {\em CommOnCV}, a system capable of representing and managing a
curriculum vitae by means of Semantic Web tools. The approach implemented by {\em CommOnCV} is based on the
concept of {\em competency} that is used to represent the knowledge/skills owned by a job seeker;
competencies are defined and described by means of suitable ontologies. In addition, {\em CommOnCV} defines a
curriculum vitae as a network of competencies and represents it by means of Semantic Web languages, like
RDF/RDFs or DAML+OIL. There are some similarities between our approach and {\em CommOnCV}; in fact, both of
them: {\em (i)} exploit ontologies to allow companies and job seekers to have a common reference for
representing competencies and tasks; {\em (ii)} represent the contents of a job proposal or a curriculum
vitae by means of suitable tools (i.e., XML in our approach, and Semantic Web languages in \cite{HaLeTr02});
{\em (iii)} are seeker oriented. As for differences, we may observe that: {\em (i)} {\em CommOnCV} is
centralized and obtrusive whereas our approach is mixed and unobtrusive; {\em (ii)} {\em CommOnCV} is not a
recommender system. Our system might inherit some features from {\em CommOnCV}. Specifically, analogously to
{\em CommOnCV}, it might represent a user profile as a network of interests, instead of a list of topics;
such a choice would allow a better comprehension of the relationships existing among user interests, as well
as a better evaluation of the relevance of each interest. However, it should be taken into account that the
exploitation of a network, instead of a list, for representing user interests, would cause an increase of the
computational complexity of our system.

\subsection{e-SRS}

In \cite{KhGo03} a multi-agent platform, called {\em e-SRS} (e-Sales Recruitment System), for recruiting and
benchmarking sales persons, is proposed. {\em e-SRS} exploits psychometric tools (e.g., questionnaires) for
classifying each user in a specific {\em category}; it considers four user categories representing specific
{\em behavioral models} (e.g., if a user rarely/frequently trusts others, if a user is/is not aggressive, and
so on). This categorization can be improved by applying a suitable clustering algorithm ({\em fuzzy
K-means}). The obtained results can be considered by a human expert for selecting the best candidates. There
are some similarities between our approach  and {\em e-SRS}. In fact, both of them: {\em (i)} exploit the
agent technology and artificial intelligence tools to overcome the limitations of existing recruitment
services; {\em (ii)} construct, handle and exploit accurate user profiles. As for differences, we may observe
that: {\em (i)} {\em e-SRS} is company oriented; {\em (ii)} it is obtrusive; {\em (iii)} it is distributed,
whereas our system is mixed; {\em (iv)} it is very specific, since it has been conceived for managing the
recruitment of {\em sales persons}; on the contrary, our approach is generic, since it can support the
recruitment activities in {\em several professional areas}; due to this reason, {\em e-SRS} produces accurate
and refined results in its application area but its extension to other professional categories seems to
require a valuable effort. {\em e-SRS} might be combined with our approach; specifically, in both the
recruitment and the benchmark of sales persons, it might take into account not only the behavioural aspects
of a job candidate but also his preferences and ``constraints''.

\section{Conclusions and hints for the future}
\label{sec:conclusions}

In this paper we have presented an XML-based, multi-agent system for supporting e-recruitment services. We
have seen that our system is characterized by various interesting capabilities, namely: {\em (i)} it allows a
uniform management of heterogeneous job proposals; as a consequence, a job seeker can pose his queries across
different, and presumably heterogeneous, job databases; {\em (ii)} it is agent- and XML-based; as a
consequence, it can easily cooperate with company information systems; {\em (iii)} the job proposal selection
algorithm takes into account information about possible interests that the user did not consider in the past
but that appear to be potentially interesting for him in the future; {\em (iv)} it is capable of handling
different recommendation strategies; {\em (v)} its recommendation strategies encompass some mathematic tools
already applied in many fields.

In the paper we have illustrated our system in detail, we have shown the obtained experimental results and,
finally, we have compared it with other related systems already presented in the literature.

Various application contexts might benefit from our research efforts. Some of the most promising ones are the
following:

\begin{itemize}

\item {\em Team Building}. In many organizations it is necessary to match individuals for composing a team
working on the same project. Such an activity, often called {\em team building} or {\em partnership building}
\cite{KeWe05}, is becoming an increasingly popular strategy for encouraging and boosting production. Team
building has been studied in a large variety of scientific areas: for instance, Human Resource based
approaches consider the skills and the abilities owned by each individual as a leading criterium to compose
teams; analogously, sociological and psychological approaches focus on the role of trust among individuals as
a key element for the success of a team. Team building activities generally consist of two phases, namely:
{\em (i)} identification of potential members of a team, and {\em (ii)} choice of individuals that really
compose a team, among the previously defined potential candidates. As observed in \cite{KeWe05}, while
various Web sites offer tools for effectively carrying out phase {\em (i)}, decision support for phase {\em
(ii)} is still rudimentary. In our opinion, our system might offer a concrete contribution in this area.
Specifically, since our approach manages a large volume of data stored in user profiles, it can elaborate
this information to compute the ``affinity degree'' existing among two individuals. In other words, an
analysis of the personal features of an individual, as well as of its preferences and exigencies, allows
managers: {\em (i)} to assign each individual to a team; {\em (ii)} to predict if trust can be established
among the potential members of a team, even if they do not share a common work history.

\item {\em Task Assignment}. A classical activity of Human Resource managers consists of associating
individuals to tasks; this activity should be carried out in such a way to exalt the skills and the
capabilities of individuals. Since this activity has a great impact on the business success, an extensive
research on it has been performed in the past \cite{Oconnel*01}. In our opinion our system could be used to
effectively solve the problem of assigning individuals to tasks. Specifically, information stored in user
profiles could be exploited for capturing user aptitudes/preferences whereas the XML-based model describing
job proposals could be used for representing task features. A matching between a user profile and a job
proposal description could be adopted for computing the so called {\em Person-Job Fit} (PJF), a parameter
used by psychologists to measure the suitability of a candidate for a specific job.

\item {\em Outsourcing}. In recent past businesses have outsourced a large variety of activities for
improving the quality of services and products, for reducing the duration of the production cycle and for
lowering costs. However, an excessive reliance on outsourcing may negatively affect business performance: for
instance, it might lead to a reduced technological innovation \cite{Kotabe92} or to a limited capability of
competing with outsourcing partners \cite{GiGrRa04}. Our system could be adopted to support business managers
to decide about the effectiveness of an outsourcing activity; specifically, it might analyze the information
stored in user profiles for assessing the capability of the internal components of an organization of
executing a given task and to evaluate if it is/is not convenient its outsourcing.

\end{itemize}

As for future work, we plan to contribute to the implementation of the ideas mentioned previously. In
addition, we plan to enrich our system by providing it with more powerful algorithms that try to predict the
future interests of a user, as well as by designing and implementing modules for supporting companies looking
for candidates. This last feature would allow the realization of a {\em hybrid} e-recruitment system,
embodying the functionalities of both seeker oriented and company oriented systems. Finally, we plan to study
the possibility to integrate our system with an e-learning framework for realizing a system capable of
suggesting (and, next, teaching) to a user the skills he should acquire for accessing new, more appealing,
job proposals.


\end{document}